\pdfoutput=1
\documentclass{aa}  

\usepackage{graphicx,natbib,url,twoopt}
\usepackage[varg]{txfonts}           
\usepackage{hyperref}   
\usepackage{booktabs}

\usepackage{url}            
\usepackage{pdfcomment}              
\usepackage{acronym} 
\usepackage{epstopdf}            
\usepackage{bm}
\usepackage{longtable}
\usepackage{caption}
\usepackage{mathtools}
\usepackage{bm}
\usepackage{esvect}
\usepackage{amsmath}
\usepackage{multirow}
\usepackage{adjustbox}
\usepackage{xcolor}
\usepackage{physics}

\hypersetup{
  colorlinks=true,  
  urlcolor=blue,    
  linkcolor=red,     
}

\bibpunct{(}{)}{;}{a}{}{,}    

\makeatletter
\newcommand{\bibnote}[2]{\global\@namedef{#1note}{#2}}
\newcommand{\biblink}[2]{\global\@namedef{#1link}{#2}}

\newcommand{\T}{$T_{\text{X}}$}
\newcommand{\Mgas}{$M_{\text{gas}}$}

\makeatother

\makeatletter
 \newcommandtwoopt{\citeads}[3][][]{%
   \nonstopmode              
   \href{http://adsabs.harvard.edu/abs/#3}%
        {\def\hyper@linkstart##1##2{}%
         \let\hyper@linkend\@empty\citealp[#1][#2]{#3}}%   %% Rutten, 2000
   \biblink{#3}{\href{http://adsabs.harvard.edu/abs/#3}{ADS}}%
   \errorstopmode}            %% fix to resume stopping at error messages 
 \newcommandtwoopt{\citepads}[3][][]{%
   \nonstopmode%              %% fix to not stop at error message in latex
   \href{http://adsabs.harvard.edu/abs/#3}%
        {\def\hyper@linkstart##1##2{}%
         \let\hyper@linkend\@empty\citep[#1][#2]{#3}}%     %% (Rutten 2000)
   \biblink{#3}{\href{http://adsabs.harvard.edu/abs/#3}{ADS}}%
   \errorstopmode}            %% fix to resume stopping at error messages
 \newcommandtwoopt{\citetads}[3][][]{%
   \nonstopmode%              %% fix to not stop at error message in latex
   \href{http://adsabs.harvard.edu/abs/#3}%
        {\def\hyper@linkstart##1##2{}%
         \let\hyper@linkend\@empty\citet[#1][#2]{#3}}%     %% Rutten (2000)
   \biblink{#3}{\href{http://adsabs.harvard.edu/abs/#3}{ADS}}%
   \errorstopmode}            %% fix to resume stopping at error messages 
 \newcommandtwoopt{\citeyearads}[3][][]{%
   \nonstopmode%              %% fix to not stop at error message in latex
   \href{http://adsabs.harvard.edu/abs/#3}%
        {\def\hyper@linkstart##1##2{}%
         \let\hyper@linkend\@empty\citeyear[#1][#2]{#3}}%  %% 2000
   \biblink{#3}{\href{http://adsabs.harvard.edu/abs/#3}{ADS}}%
   \errorstopmode}            %% fix to resume stopping at error messages 
\makeatother

\newacro{ADS}{Astrophysics Data System}
\newacro{NLTE}{non-local thermodynamic equilibrium}
\newacro{NASA}{National Aeronautics and Space Administration}

\begin{document}

\title{Detection of pure warm-hot intergalactic medium emission from a $7.2$ Mpc long filament in the Shapley supercluster using X-ray spectroscopy}

\author{K. Migkas$^{1,2,3}$, F. Pacaud$^2$, T. Tuominen$^{4,5}$, N. Aghanim$^5$}

\institute{$^1$ Leiden Observatory, Leiden University, PO Box 9513, 2300 RA Leiden, the Netherlands  \\ 
\email{kmigkas@strw.leidenuniv.nl}\\
$^2$Argelander-Institut f{\"u}r Astronomie, Universit{\"a}t Bonn, Auf dem H{\"u}gel 71, 53121 Bonn, Germany\\
$^3$SRON Netherlands Institute for Space Research, Niels Bohrweg 4, NL-2333 CA Leiden, the Netherlands\\
$^4$ Department of Physics, University of Helsinki, Gustaf H{\"a}llstr{\"o}min katu 2, FI-00014 Helsinki, Finland\\
$^5$ Universit\'e Paris-Saclay, CNRS, Institut d’Astrophysique Spatiale, 91405, Orsay, France
}

\date{Received date} 
\abstract{A significant fraction of the local Universe baryonic content still remains undetected. Cosmological simulations indicate that most of the missing baryons reside in cosmic filaments in the form of warm-hot intergalactic medium (WHIM). The latter shows low surface brightness and soft X-ray emission, making it challenging to detect. Until now, X-ray WHIM emission has been detected only in very few individual filaments, whereas in even fewer filaments WHIM was spectroscopically analyzed. The Suzaku X-ray telescope is ideal for studying X-ray WHIM emission from filaments because of its low instrumental background. We used four Suzaku pointings to study the WHIM emission of a filament in the Shapley supercluster, connecting the galaxy cluster pairs A3530/32 and A3528-N/S. We additionally employ XMM-Newton observations to robustly account for point sources in the filament, which Suzaku fails to detect because of its poor angular resolution, and to fully characterize the neighboring clusters and their signal contamination to the filament region. We report the direct imaging and spectroscopic detection of extended thermal WHIM emission from this single filament. Our imaging analysis confirms the existence of $(21\pm 3)\% $ additional X-ray emission throughout the filament compared to the sky background at a $6.1\sigma$ level. We constrain the filament gas temperature, electron density, and baryon overdensity to be $k_{\text{B}}T\approx (0.8-1.1)$ keV, $n_{\text{e}}\approx 10^{-5}$ cm$^{-3}$, and $\delta_{\text{b}}\approx (30-40)$, respectively, at a $>3\sigma$ detection level, in agreement with cosmological simulations for the first time for a single filament. Independently of the X-ray analysis, we also identify a spectroscopic galaxy overdensity throughout the filament using the Shapley Supercluster velocity Database and constrain the filament's 3D length to be 7.2 Mpc at a $53^{\circ}$ angle with the plane of the sky. Overall, this is the first X-ray spectroscopic detection of pure WHIM emission from an individual, pristine filament without significant contamination from unresolved point sources and gas clumps.}


\titlerunning{A unique, distant, merging double cluster}
\authorrunning{K. Migkas et al. }

\maketitle

\section{Introduction}\label{intro}

The framework of the standard cosmological model, $\Lambda$CDM, together with highly precise cosmological constraints from cosmic microwave background data \citep[CMB, e.g.,][]{planck20}, predicts that $\approx 5\%$ of the total matter-energy density of the Universe is in the form of baryonic matter. However, surveys in the late-time Universe only detect $\approx 60-70\%$ of the baryons expected to exist based on CMB and Big Bang nucleosynthesis data. The remaining $\approx 30-40\%$ of the expected baryons are still unaccounted for \citep[e.g.,][]{shull12}. This is known as the missing-baryon problem \citep[e.g.,][]{cen99,nicastro03,nicastro18}. Large-scale cosmological simulations suggest that a substantial fraction of the missing baryons should reside in the warm-hot intergalactic medium (WHIM), which is distributed throughout cosmic filaments \citep[e.g.,][]{cen99,martizzi19,tuominen21}. The WHIM is characterized by low gas temperatures ($T_{\text{X}}\approx [0.01-1]$ keV\footnote{Throughout the paper we use the notation $k_{\text{B}}T\equiv T_{\text{X}}$, where $k_{\text{B}}$ is the Boltzmann constant and $T$ the temperature measured in Kelvin.}) and electron densities \citep[$n_e\lesssim 10^{-4}$ cm$^{-3}$][]{galarraga21}, which make any X-ray emission from individual filaments hard to detect. Cosmic filaments connect the densest nodes of the cosmic web, that is, galaxy clusters \citep{bond96}. Therefore, the cosmic volume between cluster pairs and supercluster members is ideal for an attempt to directly detect the WHIM in individual filaments.

Several previous studies have indirectly detected the cool part ($T_{\text{X}}\approx [0.03-0.1]$ keV) of the WHIM through UV and X-ray absorption lines of distant quasars and gamma-ray bursts \citep{kaastra06,richter08,fang10,danforth16,spence24}. However, the robustness of such X-ray absorption detections remains a topic of discussion \citep[e.g.,][]{williams13,gatuzz23}. The hotter part of the WHIM ($T_{\text{X}}\approx 0.1-1$ keV) is expected to be better detected by its X-ray emission and thermal Sunyaev-Zeldovich signal \citep[tSZ][]{SZ_1972}, but it still remains largely elusive for individual filaments. Recent studies used filament stacking for the first time to report detections of X-ray WHIM emission \citep{tanimura20,tanimura22,zhang24}. Similar WHIM detections have also been reported using tSZ data from gas bridges connecting cluster pairs \citep[e.g.,][]{bonjean18, hincks22} and stacked filaments \citep{planck-filaments,degraaf19,tanimura20b}. Furthermore, recent measurements that used the dispersions of fast radio bursts showed that the baryonic content of the local Universe is, in fact, consistent with the $\Lambda$CDM predictions. This partially alleviated the missing-baryon problem \citep{macquart20,yang22}. 

The WHIM emission from individual filaments, bridges connecting galaxy clusters, and at cluster outskirts has mostly been detected through excess X-ray surface brightness compared to the sky background \citep{werner08,mitsuishi12,eckert15,ursino15,bulbul16_filam,akamatsu17,sugawara17,alvarez18,ghirardini21,reiprich21,alvarez22,dietl24,gallo24,veronica24}. Only very few of these studies were able to spectroscopically analyze the thermal WHIM emission in X-rays.

In this work, we fully characterize the WHIM emission of a 7.2 Mpc long filament in the Shapley supercluster that was neither detected nor studied in X-rays before. We provide one of the very few X-ray spectral detections and analyses of the thermal WHIM emission originating from an individual filament. Moreover, this is the first WHIM emission spectroscopic detection from such a low-density  pristine filament without considerable contamination from unresolved point sources and halos. To do this, we employed X-ray data from Suzaku and XMM-Newton pointings. The studied filament was only recently detected (independently of this study) in the optical regime and in the Shapley Supercluster velocity Database \citep{aghanim2024}. We also used the optical spectroscopic data presented in \citet{aghanim2024} to characterize the orientation of the filament in the sky. This filament connects the galaxy cluster complex A3532/A3530 and the double cluster A3528-N/S with an apparent length of $\gtrsim 1^{\circ}$. About $20^{\prime}$ of the filament lie outside $2\ R_{500}$ of any cluster (see Sect. \ref{sect:cluster-results}). Previous studies of the four clusters that are connected by the filament were carried out in the X-ray, optical, and radio domains \citep{quintana95,ettori97,bardelli01,donnelly01,venturi01,gastaldello03,lakch13,migkas20,digennaro24}, and mainly focused on the merging processes of the two separate cluster pairs and the mass of the individual clusters. All clusters have comparable intermediate masses of $\approx (1.6-2.5)\times 10^{14}\ M_{\odot}$, without extraordinary characteristics in their dynamical state or in the density and temperature profiles.

The structure of the paper is as follows: We describe in Sect. \ref{xray-optical-data} the data used in this work. In Sect. \ref{sect:cluster-results} we present the results from the analysis of the four galaxy clusters to which the filament is connected. In Sect. \ref{sect:filament-results} we present the imaging and spectral analysis of the cosmic filament. In Sect. \ref{discussion} we discuss the findings and conclusions of our work. Throughout this paper, we use a flat $\Lambda$CDM cosmology with $\Omega_{\text{m}}=0.3$ and $H_0=70$ km $s^{-1}$ Mpc$^{-1}$.

\section{X-ray and optical data}\label{xray-optical-data}

The low and stable particle-induced instrumental background (PIB) of Suzaku is a major advantage compared to other X-ray instruments and allows for the detailed study of low surface brightness (SB) diffuse systems, such as the WHIM residing in cosmic filaments. Moreover, Suzaku has a sufficiently large effective area between 0.6 and 8 keV, making it feasible to measure the \T\ of WHIM from the continuum thermal emission. The main limitation of Suzaku is the poor angular resolution, which makes the detection and treatment of point sources challenging when complementary X-ray data are not available (e.g., from XMM-Newton or Chandra) for the same sky region. To alleviate this limitation, we took advantage of XMM-Newton and its EPIC detectors (MOS1, MOS2, and PN). The latter show a relatively high and not as stable PIB level compared to Suzaku. Therefore, it is challenging to spectroscopically analyze WHIM emission with XMM-Newton. However, the point spread function (PSF) of XMM-Newton is sufficiently small, and its sensitivity at the $0.5-9$ keV band is high. These allow us to robustly identify and characterize point sources throughout the studied filament that could contaminate the WHIM emission and bias the inferred thermodynamical properties of the filament gas as determined by Suzaku. Thus, combining the strengths of the two instruments, we can accurately and precisely study the pure WHIM emission from such filaments and quantify their thermal state. In this section, we describe the data reduction, imaging, and spectral analysis of the X-ray data. Wherever applicable, we used the software package versions HEASoft 6.29, XMMSAS v19.1.0, XSPEC 12.12.0 \citep{xspec}. We also present in this section the optical data  used to characterize the length and geometry of the filament.

\subsection{XMM-Newton}
We used eight XMM-Newton pointings, listed in Table \ref{tab:pointings}. Five of them were publicly available while the other three (Obs. IDs 0861150101, 0861150201, and 0861150301, PI: F. Pacaud) were dedicated observations to better characterize the X-ray point-source population throughout the studied filament in this work.

\subsubsection{Data reduction, instrumental background, and cleaned images}
The exact procedure followed to reduce the XMM-Newton data is described in detail in \citet{migkas25}. In a nutshell, we used the reprocessed observation data files (ODF) and treated them for bad pixels, solar flare contamination, solar wind charge exchange, out-of-time events, vignetting, and exposure
time correction. All XMM-Newton EPIC cameras (MOS1/MOS2/PN) showed a lack of solar flare and wind charge exchange contamination after the respective filtering \citep[IN/OUT$<1.15$, see][for more details]{miriam, migkas25}, except for the PN camera of the 0057740301 observation. For the latter, the PN data were discarded as contaminated. Next, we inspected the so-called anomalous state of the MOS CCDs \citep{kuntz} which revealed that MOS1 CCD no. 5 is affected by the anomalous state for the observations 0030140101 and 0201780101, while MOS1 CCD no. 4 and MOS2 CCD no. 5 are affected for the 0861150101 observation. For the latter, MOS1 CCD no. 3 and no. 6 appear to be problematic due to lack of data. All the afore-mentioned CCD tiles were excluded from the rest of the analysis. After this, we masked all the detected point sources by applying a wavelet-based source detection algorithm to the combined EPIC $0.5-2$ keV image, following \citet{pacaud06}. We further performed a manual inspection to manually mask any point sources that were missed by the automatic detection. A special case is the cataclysmic binary Ex Hydrae located at (RA, DEC)=$(192.428^{\circ}, -28.978^{\circ})$, at the border of the XMM-Newton FOV for Obs. ID 0203390901 (see Fig. \ref{XMM-Suzaku-pointings-mosaic}). Due to the reduced PSF at $\approx 14^\prime$ from the center of the FOV, Ex Hydrae appears as an extended source. Therefore, a larger mask was manually applied to account for its emission. All data from the masked regions were removed from any subsequent XMM-Newton imaging and spectral analysis.

The PIB was obtained by utilizing filter wheel closed data and it was removed from the XMM-Newton images while taken into account during the spectral analysis of the filament \citep[more details on the PIB treatment method applied here, and discussion on its robustness, are given in][]{migkas24,migkas25}. Finally, we created the clean count rate mosaic image of the eight XMM-Newton pointings used in this work. To do so, we subtracted the mosaic PIB image from the clean source$+$CXB count image and we properly accounted for the varying exposure time, vignetting, and different effective areas of the EPIC MOS and PN detectors. The final clean, smoothed, PIB-subtracted, absorbed count rate XMM-Newton mosaic is shown in Fig. \ref{XMM-Suzaku-pointings-mosaic}.

\subsubsection{Spectral analysis}\label{sect:xmm-spectral}

Using XMM-Newton data, we performed the spectral analysis of the four galaxy clusters, derived their core-excised $T_{\text{X,CE}}$ values, and their \T\ profiles. The spectral fits took place in the $0.5-7$ keV band. We used Cash statistics (\texttt{cstat}) to fit the data in \texttt{XSPEC}. We did not use the XMM-Newton data to spectrally analyze the low SB filament regions to minimize any risks of systematic biases arising from the relatively high and variable PIB of XMM-Newton.

For the A3528-N/S, A3532, and A3530 pointings, there is no FOV area beyond $1.1\ R_{500}$ of any cluster and free of cluster emission.\footnote{The $R_{500}$ values are estimated in Sect. \ref{sect:cluster-results} using the $T_{\text{X,CE}}$ value and an iterative process.}. Hence, to extract the CXB spectra, we utilized the BG-X1 and BG-X2 pointings as labelled in \ref{XMM-Suzaku-pointings-mosaic}. Specifically, we obtained the CXB spectra from the $\approx 330$ arcmin$^2$ and $\approx 120$ arcmin$^2$ sky area left in BG-X1 and BG-X2 respectively after removing the $30^{\prime}$ areas around each cluster and the point sources. From the following analysis, we find that these masks correspond to $>2\ R_{500}$ for all clusters. Residual cluster emission is expected to be $\lesssim 5\%$ of the CXB level according to \citet{lyskova}. Therefore, due to the negligible contamination and large sky area, the utilized sky areas provide an accurate and precise characterization of the CXB. We used the BG-X1 CXB for the spectral fits of A3530 and A3532 and the BG-X2 CXB for the A3528-N/S spectral fits. All CXB regions are within $\lesssim 43^{\prime}$ from their respective source regions. As a cross check, we compared the CXB constraints from the BG-X1 and BG-X2 pointings and found them consistent within the statistical noise, despite the two areas being $\approx 1.1^{\circ}$ apart. Thus, it is safe to assume that the CXB level does not significantly vary throughout the mosaic of the cluster-filament system. 

The PIB spectra were considered simultaneously with the XMM-Newton CXB and source spectra by \texttt{XSPEC} when using \texttt{cstat} for Poissonian source and background spectra with an unknown model for the latter.\footnote{For details on how \texttt{XSPEC} treats the PIB spectra with \texttt{cstat}, see \url{https://heasarc.gsfc.nasa.gov/xanadu/xspec/manual/XSappendixStatistics.html} under "Poisson data with Poisson background (cstat)".} Briefly, \texttt{XSPEC} assumes that each PIB spectral bin has an unknown true value and that the PIB spectra are a Poisson sample of them. This yields a combined Poisson likelihood for the source and PIB spectra, which is then analytically maximized over the set of unknown background parameters, parallel with the source model. To alleviate any biases due to an imperfect PIB mismatch between source and FWC data, we introduced additional Gaussian lines to the model, representing Al K-$\alpha$ and Si-K-$\alpha$ fluorescence line residuals at $1.485$ keV and $1.7397$ keV respectively. We used the $0.5-7$ keV energy band for the spectral analysis, avoiding the additional PN fluorescence lines in the $7.5-9$ keV band. 

As a further check, we also constrained the CXB using ROSAT All-Sky Survey \citep[RASS][]{rass} data in the $0.1-2.4$ keV band from two circular regions of $30^{\prime}$ radius centered at (RA, DEC)$=(194.847^{\circ}, -29.328^{\circ})$ and (RA, DEC)$=(193.070^{\circ}, -30.102^{\circ})$. These regions are $1^{\circ}$ from the centers of A3532 and A3528-S respectively. We found that the RASS CXB constraints are also consistent with both the BG-X1 and BG-X2 CXB (this is partially attributed to the larger statistical uncertainties of the RASS CXB components). Therefore, we simultaneously fit the RASS and XMM-Newton CXB spectra to improve our statistics. 

The full model for XMM-Newton spectral fits is
\begin{equation}
\begin{split}
\mathtt{Model =} &\quad\mathtt{constant\times [apec_1 + tbabs\times(apec_2 +}\\
&\quad\mathtt{pow)] + \texttt{gaussians}+ tbabs\times apec_3}.\\
\end{split}
\label{eq:spectral_model}
\end{equation}
The \texttt{constant} term scales the CXB components (\texttt{apec$\mathtt{_1}$}, \texttt{apec$\mathtt{_2}$}, and \texttt{pow}) to the respective areas of the source regions. The \texttt{apec$\mathtt{_1}$} term represents the unabsorbed thermal emission from the Local Hot Bubble \citep[e.g.,][]{yeung24}. We allow its \T\ and $Z$ to vary within $(0.08-0.12)$ keV and $(0.9-1.1)\ Z_{\odot}$ respectively. Moreover, \texttt{apec$\mathtt{_2}$} represents the absorbed Milky Way Halo emission with \T\ and $Z$ left to vary within $(0.18-0.27)$ keV and $(0.9-1.1)\ Z_{\odot}$ respectively\citep[e.g.,][]{McCammon_2002}. The unresolved point sources are represented by the \texttt{pow} term with a photon index of 1.46 \citep[e.g.,][]{Luo_2017}. The X-ray absorption along the line of sight due to the Galactic interstellar medium is accounted for by the \texttt{tbabs} model. We consider the hydrogen column density parameter to be the total one (neutral+molecular, $N_{\text{H,tot}}$). We fix $N_{\text{H,tot}}$ to the value given by \citet{willingale} for each studied sky region. The \texttt{gaussians} term accounts for the residual fluorescence lines after accounting for the PIB as described earlier. The \texttt{apec$\mathtt{_3}$} term accounts for the (absorbed) cluster emission. The CXB terms are fitted using both the XMM-Newton and RASS spectra, while the \texttt{gaussians} and \texttt{apec$\mathtt{_3}$} terms are fitted using only the XMM-Newton spectra. We link the normalizations, temperatures, and metallicities of all components throughout all used spectra.

We initially fit only the CXB spectra, setting the \texttt{apec$\mathtt{_3}$} term to zero. Then, we used the CXB best-fit values as priors and we simultaneously fit the CXB and source spectra, leaving the full model free to vary. When different cluster regions were fitted simultaneously (i.e., for the multiple annuli of the \T\ profiles), we linked the CXB components throughout all different regions. Finally, we used the Cash statistic to constrain the best-fit model parameters and the \citet{aspl} abundance table in \texttt{xspec}.

\subsection{Suzaku}

\subsubsection{Data reduction, instrumental background, and cleaned images}

In this work, we used four publicly available Suzaku pointings listed in Table \ref{tab:pointings}. We utilized data from the Suzaku X-ray Imaging Spectrometer (XIS) following the Suzaku Data Reduction Guide\footnote{\url{https://heasarc.gsfc.nasa.gov/docs/suzaku/analysis/abc/}}. The HEASoft 6.29 and CALDB 20160607 versions were adopted for the entire analysis. Details on the used Suzaku pointings are given in Table \ref{tab:pointings}. For all pointings, we used the front-side illuminated (FI) CCD chips XIS0 and XIS3, and the back-side illuminated (BI) XIS1 chip.

All observations were performed with either the $5\times 5$ or $3\times 3$ editing mode and we utilized the combined data of both modes. We utilized the cleaned events files provided by the standard screening process that accounts for event grade selection and bad pixel removal and applied further filtering when necessary. Using \texttt{xselect}, we discarded data obtained within 436 s after the passage from the South Atlantic Anomaly and data taken from low elevation angles from an Earth rim (ELV$<5^{\circ}$) and a Sun-lit Earth rim (DYE$\_$ELV$<20^{\circ}$). Moreover, we discarded the observation periods with low geomagnetic cutoff rigidity
(COR$\leq 8$ GV). To avoid the charge leak effect, we removed two columns on both sides of the charge-injected
columns for the XIS1 detector. We also removed the regions at the corners of the CCD detectors where the $^{55}$Fe calibration sources are located and the parts of the XIS0 camera that were damaged by a micrometeorite impact in 2009. 

To check for potential contamination from solar wind charge exchange (SWCX), we utilized the SWEPAM/SWICS Level 3 Merged Solar Wind Proton Data\footnote{\url{https://izw1.caltech.edu/ACE/ASC/level2/sweswi_l3desc.html}} that provide the proton flux in 12-min bins. We checked the proton flux provided by SWEPAM/SWICS throughout the duration of all four Suzaku pointings. For the duration of the 808105010-808107010 pointings, the average proton flux is $\approx (0.8-1.2)\times 10^8$ cm$^{-2}$ s$^{-1}$, with the maximum proton flux for all three pointings being $2.1\times 10^8$ cm$^{-2}$ s$^{-1}$. For the 808104010 pointing the average and maximum proton flux are $1.6\times 10^8$ cm$^{-2}$ s$^{-1}$ and $3\times 10^8$ cm$^{-2}$ s$^{-1}$ respectively. \citet{yoshino} showed that the SWCX contamination levels are negligible when the proton flux is $\leq 4\times 10^8$ cm$^{-2}$ s$^{-1}$. Consequently, it is evident that the Suzaku data used in our analysis do not suffer from SWCX contamination. Furthermore, to further ensure that the utilized Suzaku data do not suffer from solar flare contamination, we examined the fluctuations of the $0.6-3$ keV light curves by grouping the data in 250 s bins. Assuming a Poisonnian distribution, we applied a $3\sigma$ clipping discarding data from time intervals that fell outside this range.\footnote{Since we study low SB emission, we wished to examine if more conservative cuts would alter our results. We repeated our analysis with a $2\sigma$ clipping instead. The final results did not noticeably changed. This further ensures the final, cleaned Suzaku data are robust and do not suffer from residual contamination.} Overall, no obvious flared periods were found and $\lesssim 1\%$ of data were discarded, suggesting these fluctuations were purely due to statistical noise.

To quantify the PIB, we followed the latest improved method provided by the XIS team\footnote{\url{https://heasarc.gsfc.nasa.gov/docs/suzaku/analysis/xisnxbnew.html}} to generate non-X-ray background (NXB) spectra with the \texttt{xisnxbgen}, \texttt{xissimarfgen}, and \texttt{xisputpixelquality} commands, and the cumulative flickering pixel map provided by CALDB. Compared to the default method to reproduce the NXB spectra \citep[e.g.,][]{tawa08}, the method adopted here significantly improves the reproduction of NXB spectra at $<1$ keV. The Night Earth data from $\pm 150$ days from each observation were used to obtain the final NXB spectra and images in the desired energy bands.  

To create the final cleaned, count rate images that were used in the SB analysis of the filament, we followed the described steps. We first used \texttt{xissim} and \texttt{xisexpmapgen} to produce the effective exposure maps that account for the vignetting (by using ray-tracing simulations of extended, spatially uniform emission) and the exposure time of the pointing. We trimmed all detector edges with $<20\%$ effective exposure compared to the maximum exposure per map to minimize the noise in our imaging analysis and avoid potential systematic biases from the vignetting correction. This threshold corresponds to effective exposures of $\lesssim (6-10)$ ks for all pointings. By applying the effective exposure maps to the count images and taking into account the different effective areas of the detectors, we obtained the combined XIS (XIS0+XIS1+XIS3) total and NXB count rate images. We then subtracted the NXB images from the total count rate ones to obtain the clean, NXB-subtracted count rate images. These correspond to the photons originating from the source (filament) and the cosmic X-ray background (CXB).

\subsubsection{Masking of point sources}\label{masking_suzaku}
The treatment of the point sources is a crucial part in the analysis of low SB sources with Suzaku due to the large PSF of the instrument. Owing to the full coverage of the sky region studied by XMM-Newton pointings, however, we are able to accurately and precisely evaluate the contamination of point sources in the Suzaku data for different techniques.

For the default analysis, we first used the $0.7-7$ keV Suzaku images and removed all detected point sources with a $2^{\prime}$ radius mask, discarding $\approx 85\%$ of the point sources' flux. These masks were applied throughout the four Suzaku pointings to the six brightest point sources with a $0.5-2$ keV flux of $f_{0.5-2}\geq 9.8\times 10^{-14}$ erg s$^{-1}$ cm$^{-2}$.\footnote{We used the $0.7-7$ keV band in Suzaku and XMM-Newton to detect active galactic nuclei (AGN) since the contrast of the latter to the source signal and X-ray foreground maximizes at such broad energy bands. However, we measured and report the AGN fluxes in the $0.5-2$ keV band because this is the energy band we use for the SB analysis of the filament later on.} as shown in the right panel of Fig. \ref{XMM-Suzaku-pointings-mosaic}. The rest of the point sources are unresolved by Suzaku, but not by XMM-Newton. Thus, we used XMM-Newton and \texttt{Xamin} \citep{pacaud06} to additionally detect all resolved point sources and measure their fluxes. Based on the cumulative $\log{N}-\log{S}$ distribution of the point sources in the filament-focused XMM-Newton pointings (Obs. ID 0861150101-0861150301) and the Chandra Deep Field South \citep[CDFS,][]{lehmer12}, we estimate that we detected active galactic nuclei (AGN) with $f_{0.5-2}\geq 4.9\times 10^{-15}$ erg $s^{-1}$ cm$^{-2}$ and $f_{0.5-2}\geq 3.2\times 10^{-15}$ erg $s^{-1}$ cm$^{-2}$ with a $90\%$ and $65\%$ completeness respectively. We apply $1^{\prime}$ radius masks in the Suzaku data at the positions of all point sources detected by XMM-Newton. This mask radius corresponds to the half-power diameter (HPD) of the Suzaku PSF and discards $\approx 50\%$ of the point source photons scattered in the Suzaku data. The followed point source masking method significantly suppresses the cosmic AGN contribution to the CXB. The residual AGN contamination can be estimated and accounted for in both the spectral and imaging analyses described in Sect. \ref{spectral-analysis-suzaku} and \ref{sect:simul-image-Suzaku} respectively.

\subsubsection{Spectral analysis}\label{spectral-analysis-suzaku}

\begin{figure*}[hbtp]
\centering
               \includegraphics[width=0.61\textwidth]{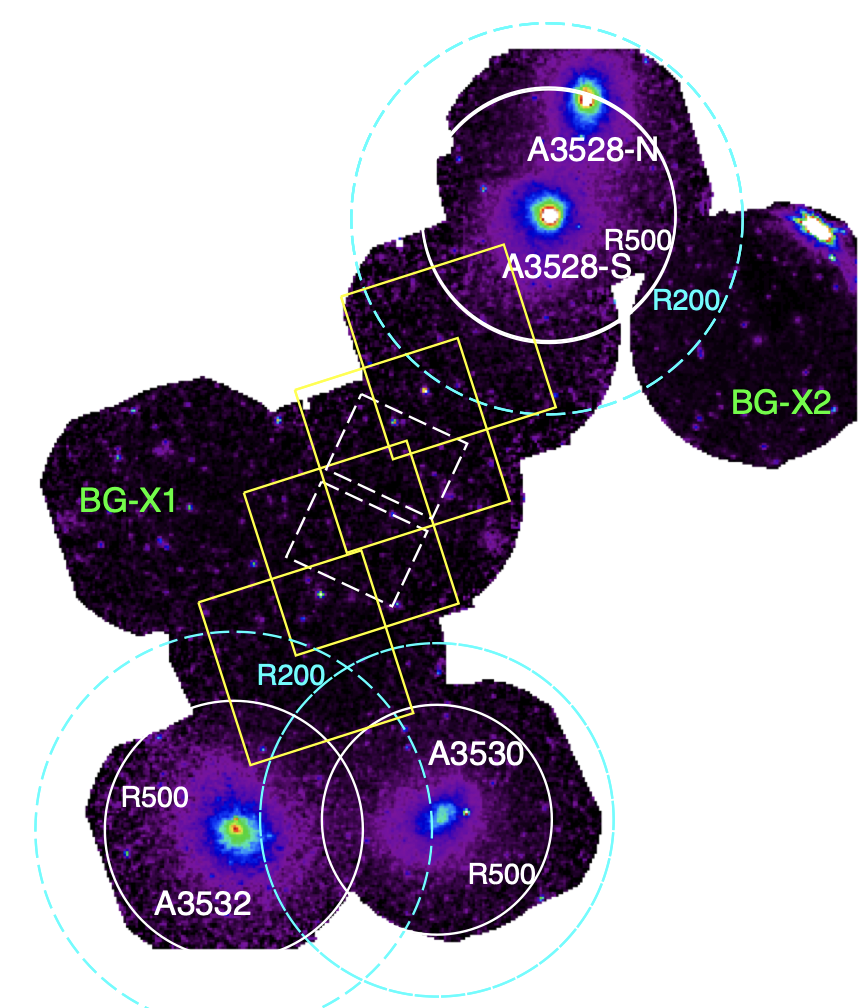}
               \includegraphics[width=0.38\textwidth]{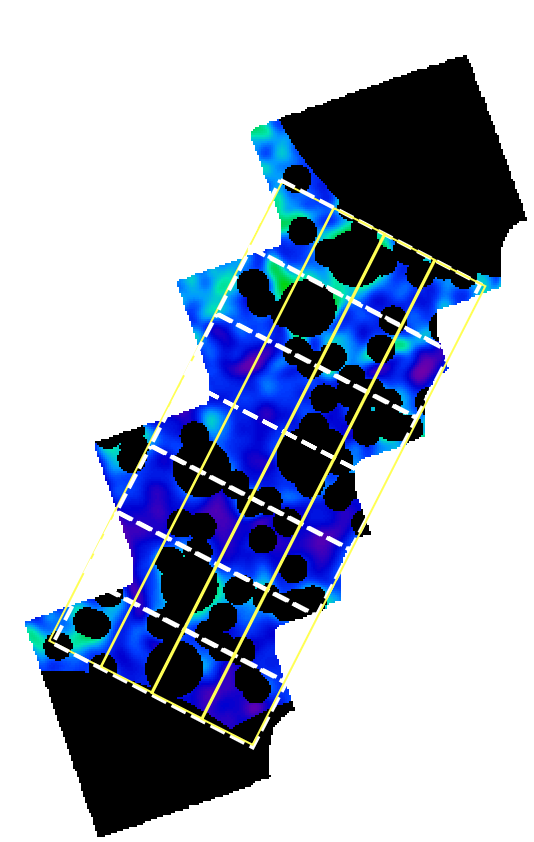}
               \caption{Cleaned, background-subtracted, and smoothed count rate mosaic images in the $0.5-2$ keV band from XMM-Newton and the entire A3532/30-A3528 system (left) and from Suzaku and the filament region alone (right). \textit{Left:} The $R_{500}$ and $R_{200}$ radii of all clusters are shown with solid white and dashed cyan circles respectively. The yellow boxes correspond to the sky areas of the four Suzaku pointings. The dashed white boxes show the regions from which we extracted and fit the Suzaku spectra to obtain the thermal properties of the filament. The BG-X1 and BG-X2 pointings show the pointings from which we extracted the CXB count rate and spectra using XMM-Newton. The point sources in the mosaic image are left unmasked for visual purposes. \textit{Right:} Regions we studied for the axial surface brightness profile (seven dashed white boxes) and for the radial surface brightness profile (four yellow solid boxes). The applied masks correspond to $1^\prime$ and $2^\prime$ masks for faint and bright AGN respectively, and to the $R_{200}$ of the surrounding clusters. }
        \label{XMM-Suzaku-pointings-mosaic}
\end{figure*}

We used the Suzaku data to analyze the spectra obtained from the central filament position, outside $2\ R_{500}$ of all clusters, as shown by the two white boxes in the left panel of Fig. \ref{XMM-Suzaku-pointings-mosaic}. We labeled "Region 1" and "Region 2" the northern and southern white-box regions, respectively. We performed spectral fits in the $0.7-7$ keV band for all XIS detectors. We used Cash statistics (\texttt{cstat}) to fit the data in \texttt{XSPEC}. We then employed an MCMC chain with the \texttt{chain} command and sample the posterior parameter distribution, from which we derived the 68.3\% parameter uncertainties using the \texttt{error} command. This choice avoids the assumptions that the \texttt{steppar} command adopts regarding the shape of the distribution of the Cash statistic. As a sanity check, we also used \texttt{plot integprob} to check if the resulted uncertainties seem reasonable. Thus, our method ensured the robustness of the parameter uncertainties. To evaluate the goodness of fit, we use the \texttt{fakeit} command in \texttt{XSPEC} and simulate 1000 set of spectra around the best-fit model with the same degrees of freedom, considering Poissonian noise and the parameter uncertainties. We fit the simulated spectra and compare the posterior C-statistic distribution with the observed value for each Region.

\paragraph{Generating response files.} To create the redistribution matrix files (RMF) and ancillary response files (ARF), we used the \texttt{xisrmfgen} and the \texttt{xissimarfgen} \citep{ishisaki} commands, respectively. The latter uses ray-tracing Monte Carlo simulations to accurately and precisely account for emission originating inside and outside the Suzaku FOV. Due to the poor PSF of Suzaku, bright sources outside the region of interest and the FOV can cause stray-light contamination. There are four bright clusters in close proximity ($\approx 34^{\prime}-50^{\prime}$) from the filament's central regions from which we extract the Suzaku spectra. Moreover, $52^{\prime}$ away, the highly bright Ex Hydrae binary is located. All of these sources can potentially cause some stray light to leak into the Suzaku spectra of the filament.\footnote{Given the distance from the studied filament regions, any such contamination is expected to be minimal}. 

To account for this, we created the ARF by feeding a simulated $100^{\prime}\times 100^{\prime}$ (2400 pixels $\times$ 2400 pixels) image to \texttt{xissimarfgen}, using the \texttt{source\_mode=SKYFITS} option. In the simulated image, we reproduced the emission of the four galaxy clusters based on their SB profiles as they are derived in Sect. \ref{cluster-profiles}. We also included the emission of the filament as obtained from its transverse and radial SB profiles in Sect. \ref{sect:imaging-filam-results}. We extrapolated the transverse emission of the filament based on the average profiles given in \citet{galarraga22}. Moreover, we inputted the emission of all detected AGN based on their measured flux (Sect. \ref{masking_suzaku}) at the central pixel of their detection, and the CXB emission. Furthermore, we simulated the X-ray emission of Ex Hydrae based on the best-fit model provided by \citet{pekon}, averaging the results from different epochs. In the ARF calculation, we also included the decrease in the soft response of the XIS detectors due to the contamination of the optical blocking filter. Finally, we correct all the normalizations of the fitted model components (see Eq. \ref{eq:spectral_model_suzaku} for the known caveat of the ARF normalization, which is based on the flux of the entire simulated image and not just the spectral extraction region.\footnote{See \url{https://heasarc.gsfc.nasa.gov/docs/suzaku/analysis/xissimarfgen/xissimarfgen_commandoptions.html}.}

\paragraph{NXB treatment.} Prior to the spectral fits, we checked that the obtained NXB spectra match the observed spectra in the $7-12$ keV energy band. At these energies, the NXB dominates the CXB \citep[e.g.,][]{kettula}, and no filament emission is expected. Indeed, the NXB spectra nicely match the observed spectra. Thus, we treat the obtained NXB spectra with \texttt{XSPEC} and \texttt{cstat} as described in Sect. \ref{sect:xmm-spectral} for the XMM-Newton data.\footnote{Note that the Suzaku XIS NXB is rather low and stable, and the source spectra are (much) higher at $\lesssim 3-4$ keV where most of the filament's emission comes from. Thus, the best-fit model of the filament is rather insensitive to the exact treatment of the NXB.} Similarly to the XMM-Newton analysis, we add Gaussian lines to the model at 1.49, 1.74, 2.12, 5.90, and 6.49 keV \citep{tawa08} to account for any fluorescence line residuals due to an imperfect NXB estimation. 

\paragraph{Residual masked AGN emission treatment.} Moreover, we add an extra $\mathtt{pow_2}$ term with a fixed photon index of 1.46 (same as for $\mathtt{pow}$ in Eq. \ref{eq:spectral_model}) to account for the contamination from the leaked photons of the masked AGN. For each spectral fit, we set the starting point normalization of this component to the value that replicates the expected residual flux of the masked AGN in each region.\footnote{As explained in Sect. \ref{masking_suzaku}, we know the flux of each AGN from the XMM-Newton data and we know what fraction of the AGN photons that leak from the applied mask, based on the mask radius and the PSF of Suzaku. Thus, we have an estimate of the respective contaminating flux in each region.} The normalization of the $\mathtt{pow_2}$ term is left free to vary within $30\%$ of the starting value. This is done to account for any cross-calibration differences between XMM-Newton and Suzaku (see Sect. \ref{cross-calib}) and the time-variability of AGN detected by the two instruments in different years (although this effect should average out when considering tens of AGN). The full model for the Suzaku spectral fits is given by:
\begin{equation}
\begin{split}
\mathtt{Model =} &\quad\mathtt{constant\times [apec_1 + tbabs\times(apec_2 +}\\
&\quad\mathtt{pow_1)] + \texttt{gaussians}+ tbabs\times (apec_3+pow_2)}.\\
\end{split}
\label{eq:spectral_model_suzaku}
\end{equation}
All previously introduced terms have the same meaning as in Eq. \ref{eq:spectral_model}. The $\mathtt{apec_3}$ term now accounts for the emission of the filament. 

\paragraph{Fitting process.} Since the Suzaku pointings do not cover any sky area free of filamentary emission, we additionally utilized the XMM-Newton and RASS data as described in Sect. \ref{sect:xmm-spectral} to constrain the CXB. The spectrally analyzed filament regions lie roughly between the two RASS regions and the BG-X1 and BG-X2 pointings. Given that all the above-mentioned CXB regions provide consistent results (see Sect. \ref{sect:xmm-spectral}, we jointly model all the CXB data from the 450 arcmin$^2$ XMM-Newton area and the 1.57 deg$^2$ RASS area around the regions of interest. For the spectral analysis of the low SB emission of the filament, it is crucial to have an unbiased estimation of the CXB. To ensure this, we only consider the $0.5-2.5$ keV band for XMM-Newton, that minimizes the effects the PIB treatment might have on the XMM-Newton spectra. To constrain all the terms in Eq. \ref{eq:spectral_model_suzaku}, we perform the following. First, we fit all the RASS and XMM-Newton CXB data together with the Suzaku data in the $4-7$ keV band, where no significant filament emission is expected given its expected low \T. For the RASS and XMM-Newton spectra, $\mathtt{apec_3}$ is set to zero (and $\mathtt{pow_3}$ is not included in their model). For the Suzaku spectra, we consider only the $\mathtt{pow_1}$ and $\mathtt{pow_2}$ terms, setting the normalizations of all $\mathtt{apec}$ terms to zero. We link the normalization of $\mathtt{pow_1}$ throughout all different spectra. This process ensures that the CXB and the residual masked AGN contamination in the Suzaku spectra are robustly constrained. After that, we use the best-fit CXB parameter values as starting points and refit all the available spectra simultaneously, this time considering the full $0.7-7$ keV band for Suzaku and letting the normalization and \T\ of $\mathtt{apec_3}$ free to vary. The metal abundance is set to $0.1\ Z_{\odot}$ for the default analysis. Values of $0.05\ Z_{\odot}$ and $0.15\ Z_{\odot}$ are also tested to estimate the systematic uncertainty induced by the unknown true $Z$. The variance of the best-fit parameters are then added in quadrature to the statistical uncertainties.

\subsection{Optical data}\label{optical-data}

The data used in our analysis of the filaments traced by galaxies are the same as those which served to reconstruct the 3D filamentary structure of the Shapley supercluster over an area of about $300$ square degrees by \cite{aghanim2024}. These publicly available data (Shapley Supercluster Velocity Database) compiled by \citet{Quintana2020} cover a region within 12h43m00s < RA < 14h17m00s and $-23^{\circ}30'00''$ > Dec > $-38^{\circ}30'00''$, containing 18,146 velocity measurements for 10,719 galaxies. Only galaxies with velocities ranging between 9,000 and 18,900 km/s ($0.03<z<0.063$) corresponding to the spectroscopic redshifts of galaxies belonging to the Shapley supercluster were considered. Correction for fingers-of-God effects \citep{FoG-1972} due to clusters and groups of galaxies  was performed \citep[see][for details]{aghanim2024}, which allowed us to correct for stretched distributions of galaxies at the position of clusters and/or groups, that could be mistaken for actual filaments.

In \cite{aghanim2024}, the detection of filaments constituting the full filamentary structure of the Shapley supercluster was performed with the Tree-based Ridge Extractor (T-REx) algorithm \citep[][to which we refer the reader for details on the algorithm]{Bonnaire2020,Bonnaire2021arXiv}. T-REx estimates the filamentary pattern based on a graph modeling of the galaxy distribution by building a smooth version of the minimum spanning tree using a Gaussian mixture model to describe the spatial distribution of galaxies. The resulting filamentary network for the entire Shapley supercluster region \citep{aghanim2024} includes the filament connecting the clusters A3530/32 and A3528-S, detected independently in the X-rays in this work.

\begin{figure*}[hbtp]
\centering
               \includegraphics[width=0.49\textwidth]{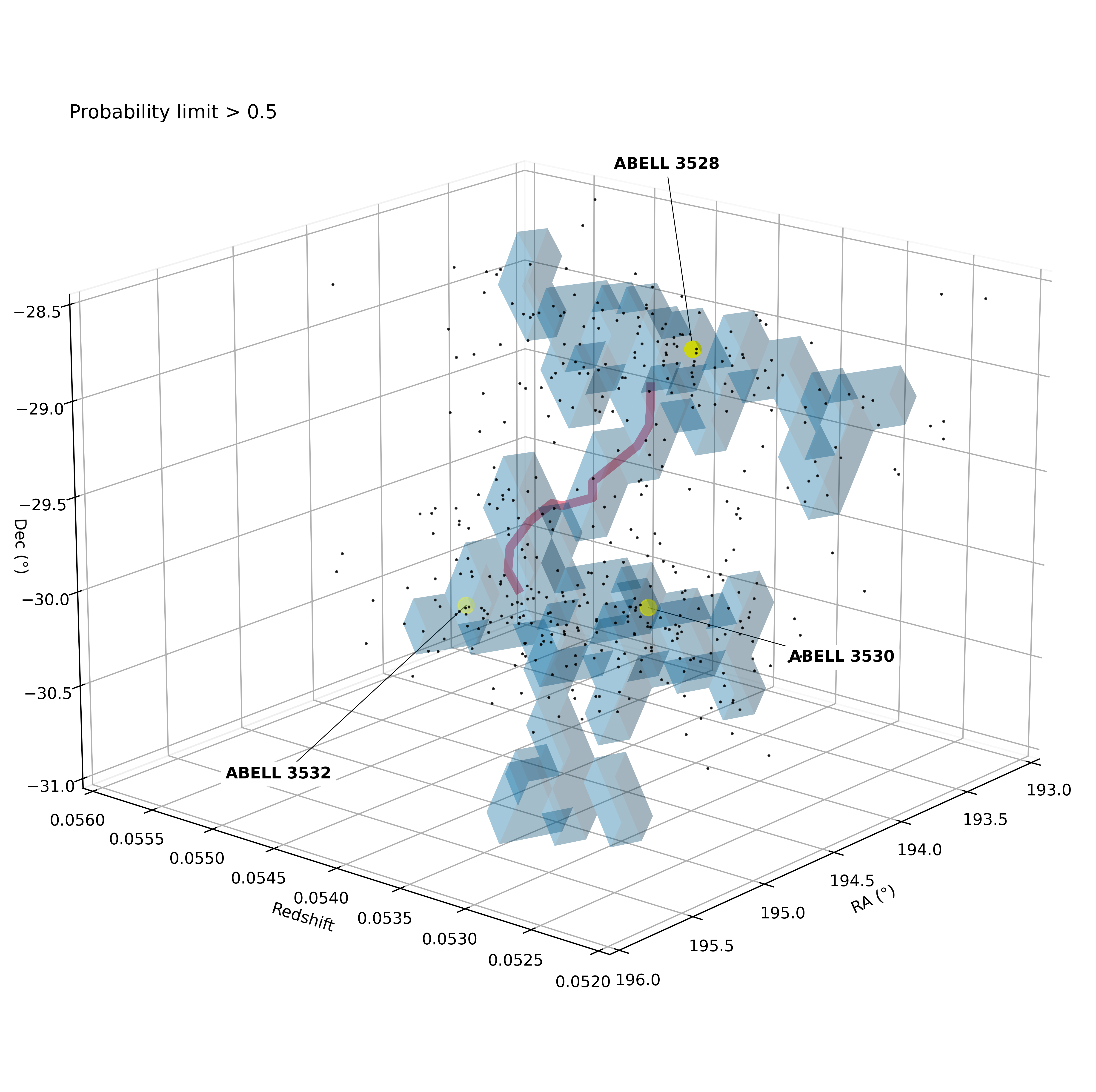}
               \includegraphics[width=0.49\textwidth]{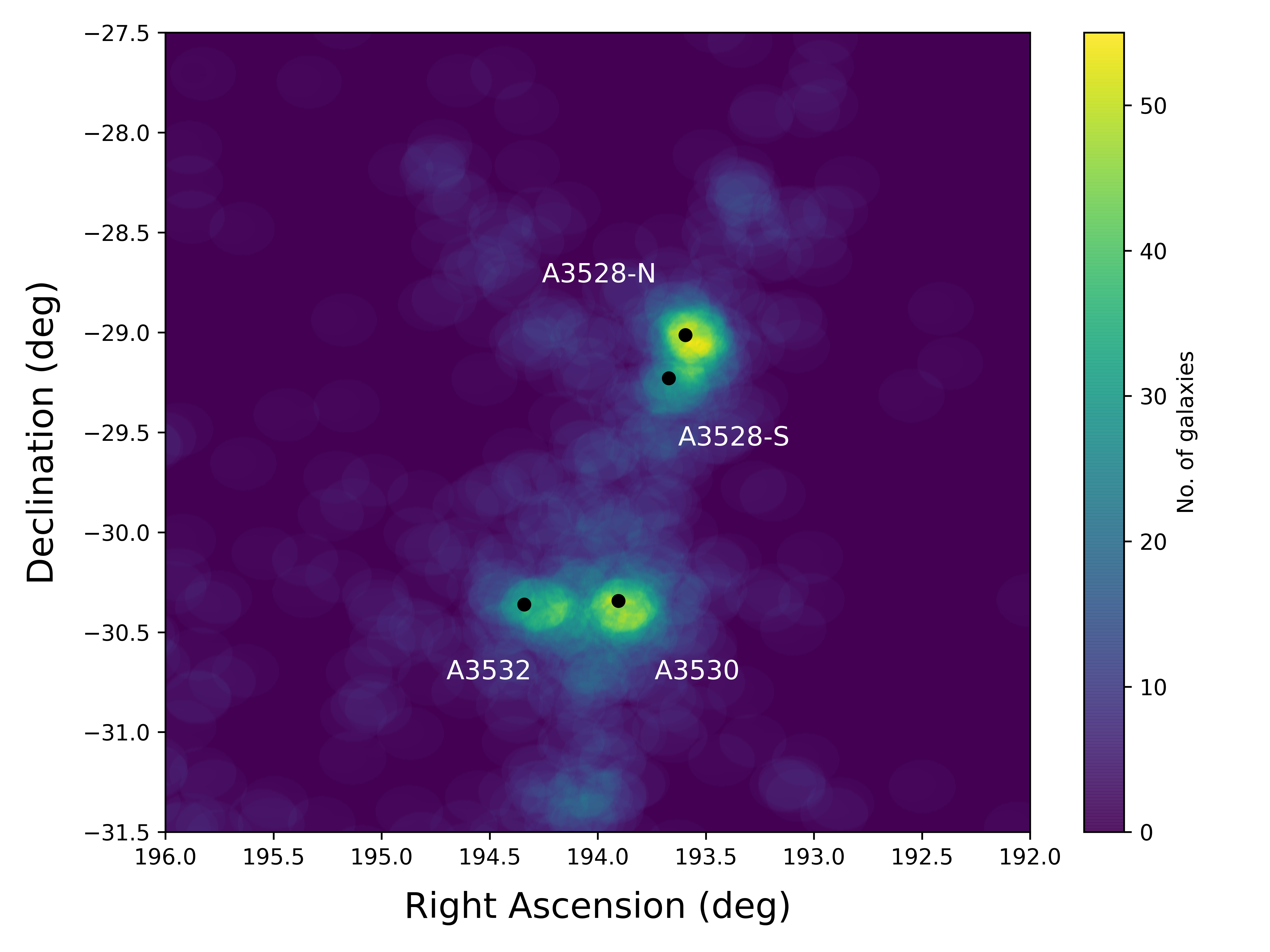}
               \caption{3D (left) and 2D (right) galaxy overdensity throughout the filament based on the spectroscopic optical data. \textit{Left:} Spectroscopic galaxies and cluster centers are shown as black and yellow large dots respectively. The filament volumes detected with T-REx are shown as blue voxels (with a probability cut at 0.5), while the central spine extracted with COWS is shown with a red line. \textit{Right:} The number of galaxies within $8^\prime$ from each pixel is displayed color-coded. The black dots show the X-ray cluster centers. The space between the galaxy clusters (identified as the highest galaxy density peaks) shows a clear overdensity compared to the background.}
        \label{fig:galaxy-overdensity}
\end{figure*}

To estimate the reliability of the filaments detected with T-REx, the detection process was repeated 100 times with randomly selected subsamples of the same input galaxy distribution. The output is a 3D probability grid, where a value between 0 and 1, corresponding to the fraction of filaments from the 100 realizations that are detected, was assigned to each voxel. The filaments were then selected by setting a lower limit to the probability value. Thus, the higher the limit, the more reliable and conservative the reconstructed filamentary network is (see left panel of Fig. \ref{fig:galaxy-overdensity} with a probability value of $>0.5$). For further details on the application of the filament to the Shapley supercluster, see \citet{aghanim2024}.

For our analysis of the specific filament connecting the clusters A3530/32 and A3528-S in the present study, we need to identify the spine so that we estimate the physical properties of the filament, such as its geometry and hence its length. We used the COsmic Web Skeleton \citep[COWS,]{Pfeifer2022} tool which identifies the central ridge of a filament by gradually thinning the voxels, that is, removing outer layers until a single set of subvoxels remain, tracing the core of the filament volume and hence the spine. We post-processed the original output of \cite{aghanim2024} (obtained with a probability value of $>0.5$) with the COWS tool and show as a red line in the left panel of Fig. \ref{fig:galaxy-overdensity} the resulting spine of the filament connecting A3530/32 and A3528. Finally, the projected optical galaxy overdensity throughout the filament and the four galaxy clusters is shown in the right panel of Fig. \ref{fig:galaxy-overdensity}.

\begin{table*}[!ht]
    \centering
    \caption{All Suzaku and XMM-Newton observations used in this work.}
    \begin{tabular}{c c c c c}
    \hline
    \hline
Observing Date & ObsID & GTI [ks] & R.A. (J2000) & Dec. (J2000) \\
\hline
 & & Suzaku & & \\
%\multicolumn{5}{c}{Suzaku}\\
\hline
Jul 2013 & 808104010 & 36.7 &  12h 55m 31.22s  & -29d 28$^\prime$ 36.5$^{\prime \prime}$  \\
Jul 2013 & 808105010 & 37.1 & 12h 55m 54.77s  & -29d 39$^\prime$ 00.4$^{\prime \prime}$   \\
Jul 2013 & 808106010 & 54.5 & 12h 56m 20.95s  & -29d 50$^\prime$ 25.4$^{\prime \prime}$   \\
Jul 2013 & 808107010 & 28.6 & 12h 56m 44.33s  & -30d 02$^\prime$ 36.6$^{\prime \prime}$   \\
%8-9 March2020
\hline
 & & XMM-Newton & & \\
%\multicolumn{5}{c}{XMM-Newton}\\
\hline
Dec 2001 & 0030140101 & 16.5, 16.5, 11.8 & 12h 54m 33.8s & -29d $08' 30.0''$ \\
Jul 2002 & 0030140301 & 8.5, 8.4, 5.6 & 12h 57m 21.99s & -30d $22' 03.0''$ \\
Jan 2003 & 0057740301 & 41.1, 63.5, - & 12h 52m 54.22s & -29d $27' 06.7''$ \\
Jan 2004 & 0201780101 & 12.2, 12.3, 10.9 & 12h 55m 30.68s & -30d $19' 53.0''$ \\
Jun 2004 & 0203390901 & 11.7, 12.2, 10.6 & 12h 57m 44.89s & -29d $45' 59.0''$ \\
Jun 2020 & 0861150101 & 8.6, 8.5, 6.7  & 12h 55m 59.04s & -29d $44' 46.0''$ \\
Jun 2020 & 0861150201 & 8.6, 8.6, 6.4  & 12h 55m 09.98s & -29d $26' 42.0''$ \\
Jul 2020 & 0861150301 & 13.4, 13.5, 11.6 & 12h 56m 38.84s & -30d $03' 10.4''$ \\
\hline
\hline
    \end{tabular}
    \label{tab:pointings}
\end{table*}

\section{Analysis of the four galaxy clusters}\label{sect:cluster-results}

Prior to the analysis of the filament, we need to robustly characterize the surface brightness, gas density, and temperature profiles of the four galaxy clusters to distinguish their signal from the X-ray emission and gas properties associated with the filament. Not doing so might result in overestimating the emission level and density of the filament connecting the two cluster pairs.
\subsection{X-ray emission peak, cluster radius, and total mass}
As a first step, we need to determine the cluster centers. We consider the X-ray emission peaks as the centers of the four systems. We identified each X-ray peak using the cleaned, smoothed, count rate $0.4-1.25$ keV XMM-Newton images and following the method explained in detail in \citet{migkas25}.

To quantify the $R_{500}$ radii of the four clusters and their total mass $M_{500}$ within $R_{500}$, we use the $Y_{\text{X,CE}}-M_{\text{500}}$ relation of \citet[][hereafter L15]{lovisari} and the definition of $R_{500}\propto M_{500}^{1/3}$. The X-ray Compton$-y$ equivalent is $Y_{\text{X,CE}}=M_{\text{gas}}\times T_{\text{X,CE}}$, with \Mgas\ being the gas mass within $R_{500}$ and $T_{\text{X,CE}}$ being the average core-excised gas temperature within $(0.15-1)\times R_{500}$. The method for measuring \Mgas\ is described in Sect. \ref{cluster-profiles}. We followed an iterative process starting from an arbitrary $R_{500}$ until the iteration converged within $2\%$ to a final value. Following \citep{reiprich13}, the $R_{200}$ and $M_{200}$ were then estimated using $R_{200}=1.538\ R_{500}$ and $M_{200}=\frac{200}{500}\times 1.538^3\ M_{500}$. 

To ensure that the \Mgas\ and \T\ measurements for each cluster are not significantly contaminated by the emission of neighboring clusters, we performed the following. For A3532, A3530, and A3528-S, we first only considered the clusters' half-circles opposite to the neighboring cluster, avoiding any strong residual contamination. After we obtained the first $R_{500}$ and $R_{200}$ estimates, we masked the entire $R_{200}$ ($R_{500}$) area of A3530 (A3528-S) in order to measure the \Mgas\ and \T\ of A3532 (A3528-N). After obtaining the final radii for A3532 and A3528-N, we masked these clusters accordingly to measure the properties of their neighboring clusters. The final coordinates, radii, and mass estimates for all clusters are displayed in Table \ref{X-ray-table}. 

All four clusters appear to have very similar radii, ranging within $R_{500}\approx (820-920)$ kpc. Similarly, the masses of all clusters vary within $M_{500}\approx (1.6-2.5)\times 10^{14}\ M_{\odot}$. The $R_{500}$ radii of both cluster pairs overlap, although only marginally for the A3532-A3530 pair. For the latter pair, the $R_{200}$ radii do not encompass the center of the neighboring cluster. In contrast, the clusters in the A3528-N/S pair lie closer to each other and their $R_{500}$ radii cross the other cluster's center. 

\begin{table*}[htbp]
\centering
\caption{X-ray properties of the four galaxy clusters in the system.}
\label{X-ray-table}
\begin{tabular}{lcccc}
\hline
\hline
Cluster property & A3532 & A3530 & A3528-S & A3528-N \\
\hline
(1) R.A. & 194.340$^{\circ}$ & 193.904$^{\circ}$ & 193.671$^{\circ}$ & 193.594$^{\circ}$ \\
(2) DEC. & $-30.362^{\circ}$ & $-30.344^{\circ}$ & $-29.230^{\circ}$ & $-29.014^{\circ}$ \\
(3) $z$ & 0.055 & 0.054 & 0.054 & 0.054 \\
(4) $L_{\text{X}}$ & $8.77\pm 0.41$ & $3.81\pm 0.23$ & $7.59\pm 0.36$ & $6.51\pm 0.33$ \\
(5) $T_{\text{X,CE}}$ & $4.54^{+0.39}_{-0.26}$ & $3.27^{+0.17}_{-0.16}$ & $3.86^{+0.17}_{-0.17}$ & $4.19^{+0.49}_{-0.47}$ \\
(6) $Z_{\text{X,CE}}$ & $0.41^{+0.06}_{-0.05}$ & $0.57^{+0.08}_{-0.09}$ & $0.43^{+0.07}_{-0.07}$ & $0.32^{+0.07}_{-0.08}$ \\
(7) $M_{\text{gas}}$ & $2.11\pm 0.05$ & $1.37\pm 0.06$ & $2.00\pm 0.04$ & $1.91\pm 0.07$ \\
(8) $Y_{\text{X,CE}}$ & $9.58^{+0.85}_{-0.59}$  & $4.48^{+0.30}_{-0.28}$ & $7.72^{+0.37}_{-0.37}$ & $8.00^{+0.98}_{-0.94}$ \\
(9) $M_{500}$ & $2.49\pm 0.12$ & $1.65\pm 0.08$  & $2.26\pm 0.10$ & $2.32\pm 0.11$ \\
(10) $M_{200}$ & $3.62\pm 0.18$ & $2.40\pm 0.12$ & $3.29\pm 0.16$ & $3.38\pm 0.17$ \\
(11) $R_{500}$ & $14.34^{\prime}/920$ kpc & $12.78^{\prime}/820$ kpc & $14.17^{\prime}/909$ kpc & $14.27^{\prime}/915$ kpc \\
(12) $R_{200}$ & $22.05^{\prime}/1415$ kpc & $19.65^{\prime}/1259$ kpc & $21.79^{\prime}/1398$ kpc & $21.95^{\prime}/1407$ kpc \\

\hline
\end{tabular}
\tablefoot{(1) Right Ascension J2000. (2) Declination J2000. (3) Optical spectroscopic redshift. (4) X-ray luminosity in the $0.5-2$ keV band measured within $<R_{500}$ in $10^{44}$ erg s$^{-1}$ units. (5) Core-excised ICM temperature measured within $(0.15-1)\times R_{500}$ in keV units. (6) Same as (5) but for the metallicity $Z_{\text{CE}}$ in solar metallicity units. (7)  Gas mass measured within $<R_{500}$ in ($10^{13}\ M_{\odot}$) units. (8) X-ray equivalent of the Compton$-y$ parameter in $10^{13}\ M_{\odot}\ $keV units. (9) Total mass estimated within $<R_{500}$ in $10^{14}\ M_{\odot}$ units using the $Y_{\text{X,CE}}-M_{500}$ relation of L15. (10) Total mass within $<R_{200}$ in $10^{14}\ M_{\odot}$ units using the $R_{200}=1.538\ R_{500}$ relation of \citet{reiprich13} and the determined $R_{500}$ value. (11) $R_{500}$ radius (12) $R_{200}$ radius.}
\end{table*}

\subsection{Surface brightness and gas density profiles}\label{cluster-profiles}

\subsubsection{Profile extraction and fitting}
To derive the X-ray SB and electron number density ($n_e$) cluster profiles, alongside the X-ray luminosity ($L_{\text{X}}$), and \Mgas, we used the XMM-Newton data and \texttt{pyproffit} \citep{eckert20}. The latter deconvolves the observed SB profiles accounting for the point spread function (PSF) of XMM-Newton and applies a multiscale decomposition to provide the deprojected $n_e$ profile. To extract the profiles, we use radial bins of $20^{\prime \prime}$. We adopt a double-$\beta$ model to fit both profiles with the form:
\begin{equation}
    \text{SB}(r)=aS_{0}\left(1 + \frac{r^2}{r_{c,1}^2}\right)^{-3\beta_1+0.5} + (1-a)S_{0}\left(1 + \frac{r^2}{r_{c,2}^2}\right)^{-3\beta_2+0.5}
    \label{SB_eq}
\end{equation}
and
\begin{equation}
    n_e(r)=\left(n_{0,1}^2\left(1 + \frac{r^2}{r_{c,1}^2}\right)^{-3\beta_1} + n_{0,2}^2\left(1 + \frac{r^2}{r_{c,2}^2}\right)^{-3\beta_2} \right)^{\frac{1}{2}}
    \label{nh_eq}
\end{equation}
for the SB and $n_e$ profiles, respectively. Here, $S_0$ is the central SB for $r=0$, while $a\in (0,1)$ accounts for the fractional contribution of each component to the total SB and $n_e$. Moreover, $r_c$, $\beta$, and $n_e$ are respectively the core scale, slope, and central gas density of the two profile components. We find that the double-$\beta$ models describe the data sufficiently well.\footnote{For A3532 and A3530 we find that even a single-$\beta$ model describes the data sufficiently well.  Integrating the best-fit, PSF-corrected profiles out to $R_{500}$ we obtain the X-ray properties for each cluster.} 

We measure the CXB $0.5-2$ keV count rate from the total $\approx 450$ arcmin$^2$ CXB sky area in the XMM-Newton pointings and the two RASS areas as described in Sect. \ref{sect:xmm-spectral}. Owing to the consistency of the CXB SB (in flux units) measured by XMM-Newton and RASS, we take the weighted average SB as the overall CXB level. Overall, we find the SB of the joint $0.5-2$ keV CXB data to be $(2.66\pm 0.08)\times 10^{-15}$ erg $s^{-1}$ cm$^{-2}$ arcmin$^{-2}$. 

\begin{figure*}[hbtp]
\centering
               \includegraphics[width=0.45\textwidth]{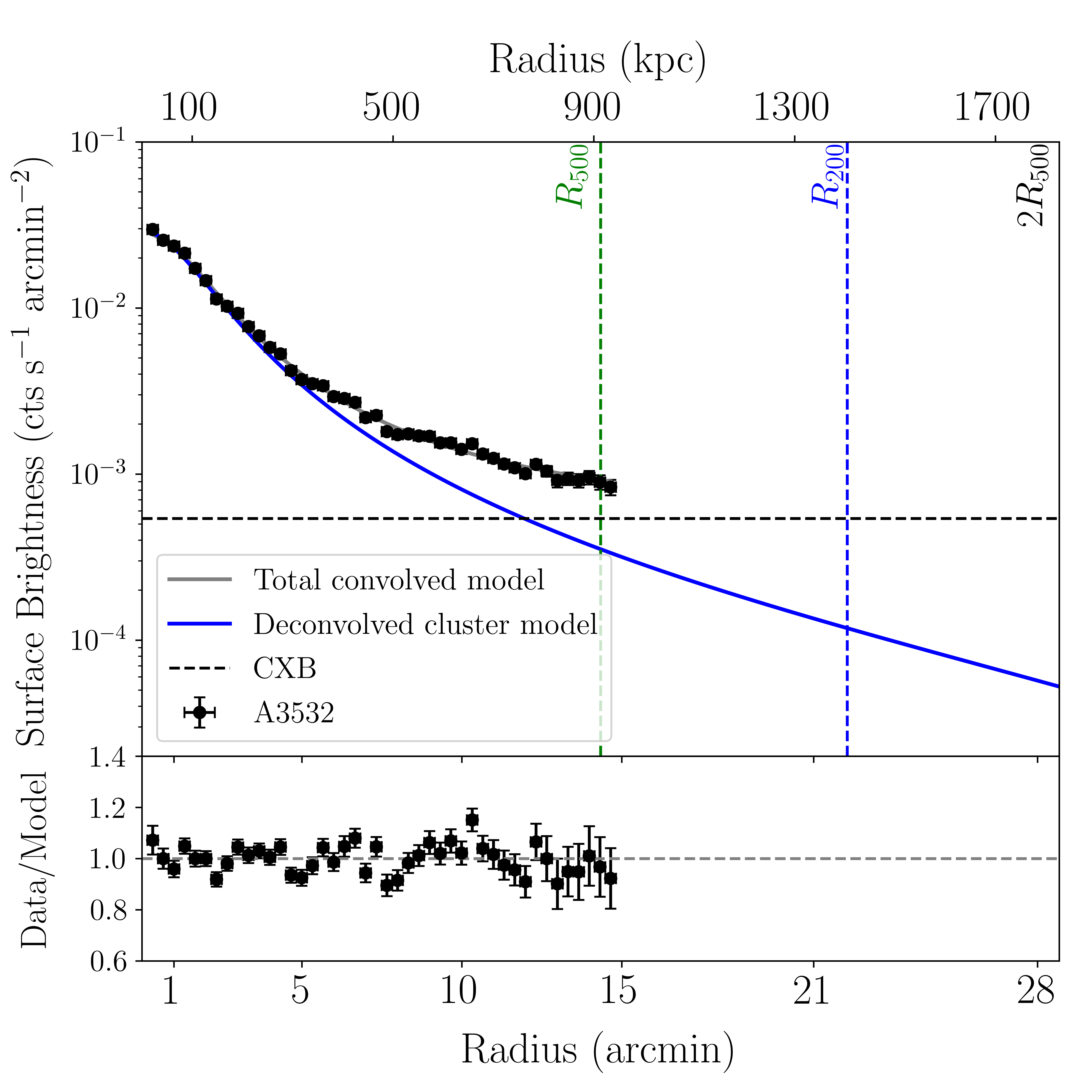}
               \includegraphics[width=0.45\textwidth]{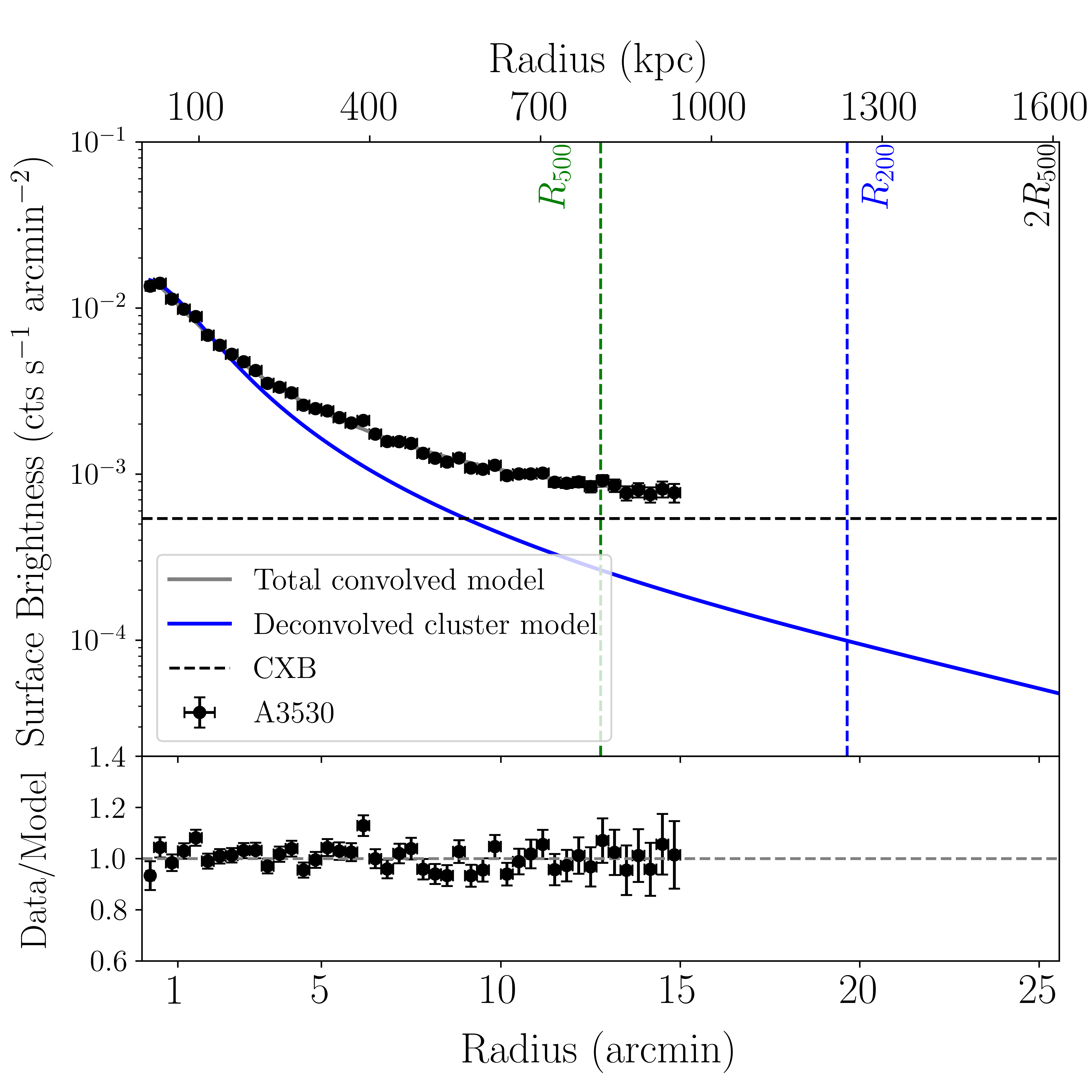}
               \includegraphics[width=0.45\textwidth]{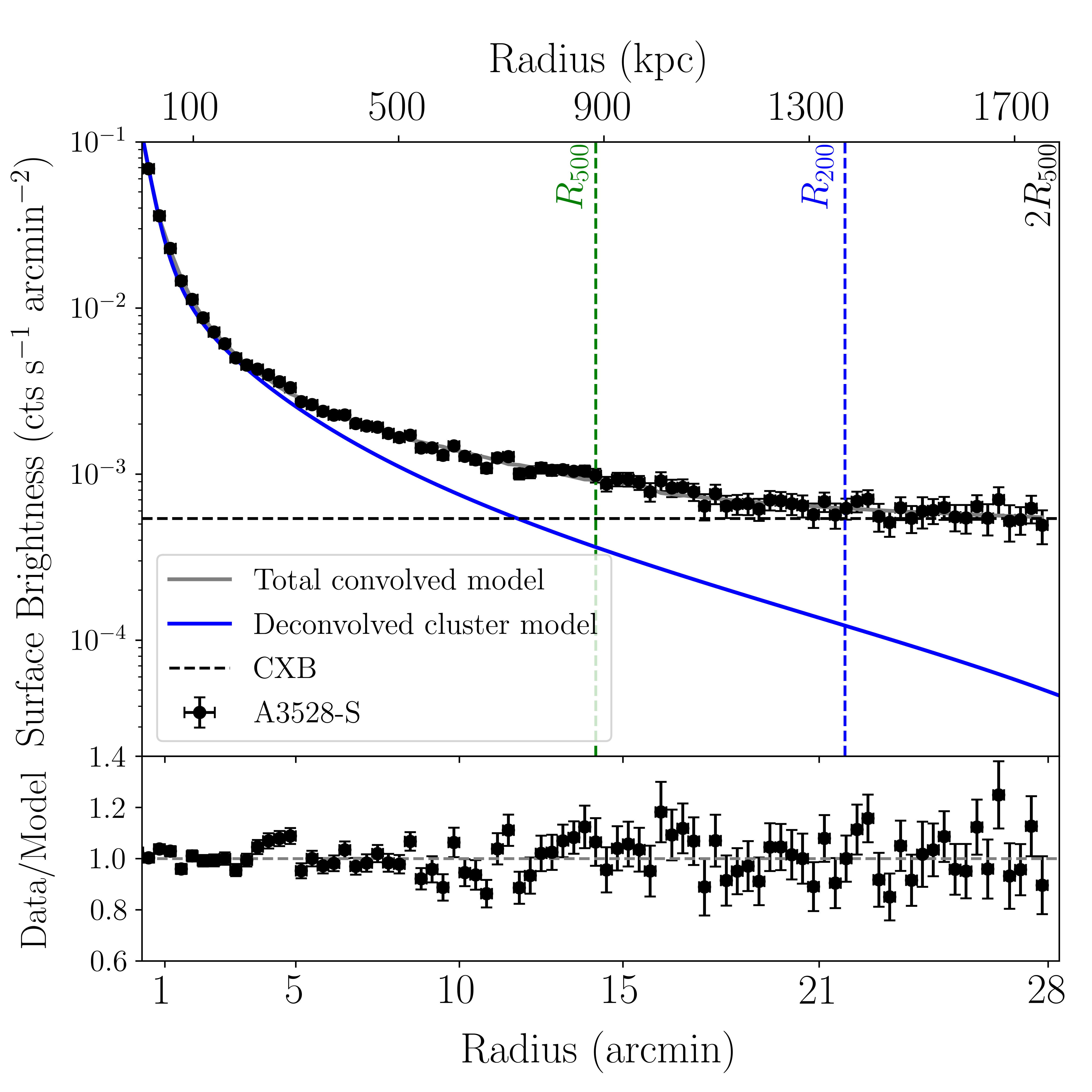}
               \includegraphics[width=0.45\textwidth]{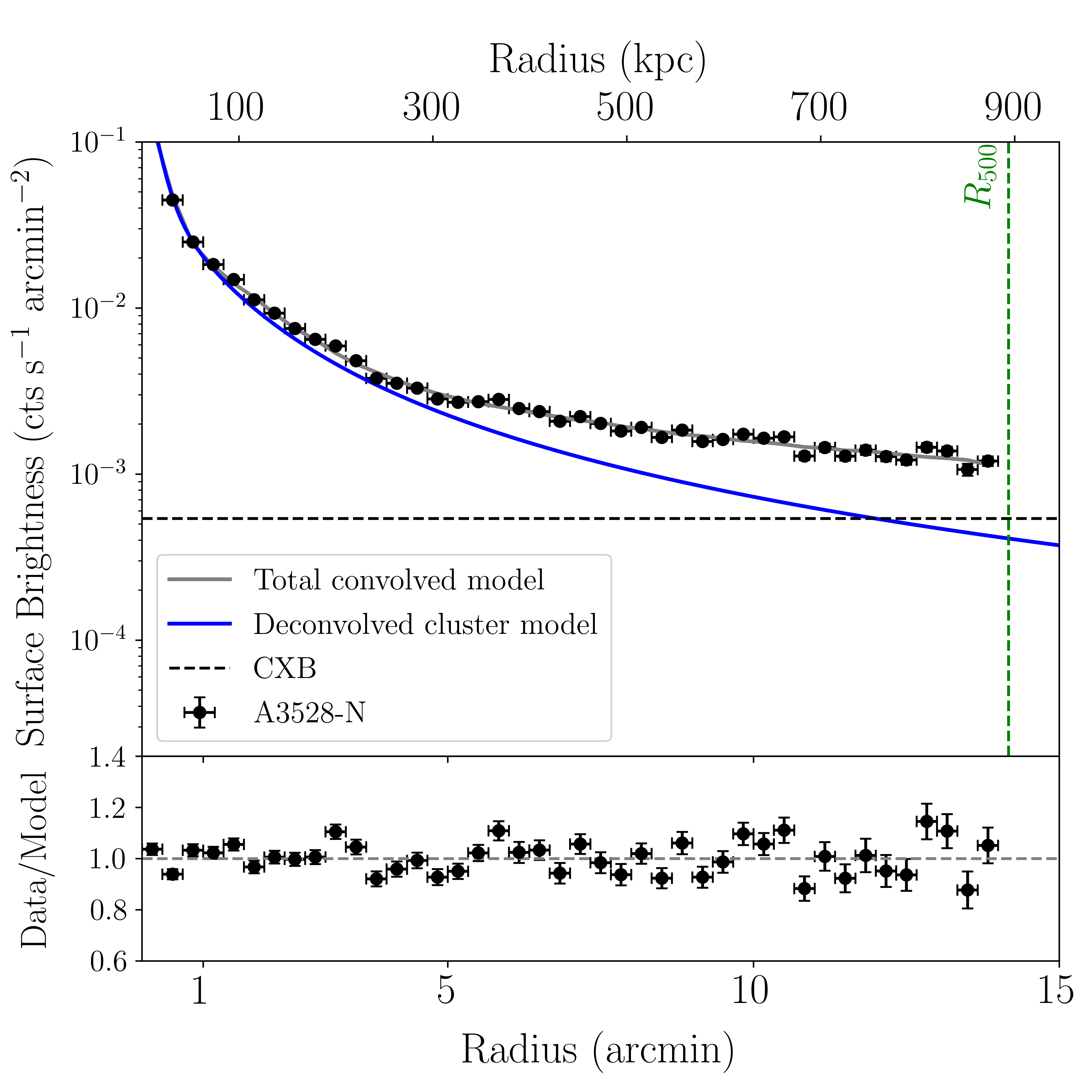}
               \caption{Surface brightness profiles (PIB-subtracted) for the A3532 (top left), A3530 (top right), A3528-S (bottom left), and A3528-N (bottom right) clusters in the $0.5-2$ keV band. The total (cluster$+$CXB), PSF-convolved profiles are displayed in gray, and the best-fit cluster-only PSF-deconvolved profiles are shown by the blue line. The latter is extrapolated up to $2\ R_{500}$, except for A3528-N, which is only plotted until $\approx R_{500}$ since it is irrelevant for the analysis of the filament (this plotting difference is the reason why this SB profile appears to be flatter than the others). The $R_{500}$ and $R_{200}$ values are displayed with the dashed green and vertical blue lines, respectively. The CXB level is displayed with the horizontal black line. The residuals of the fit are shown for all clusters in the bottom subpanels.}
        \label{SB-profiles}
\end{figure*}

\begin{figure}[hbtp]
\centering
               \includegraphics[width=0.45\textwidth]{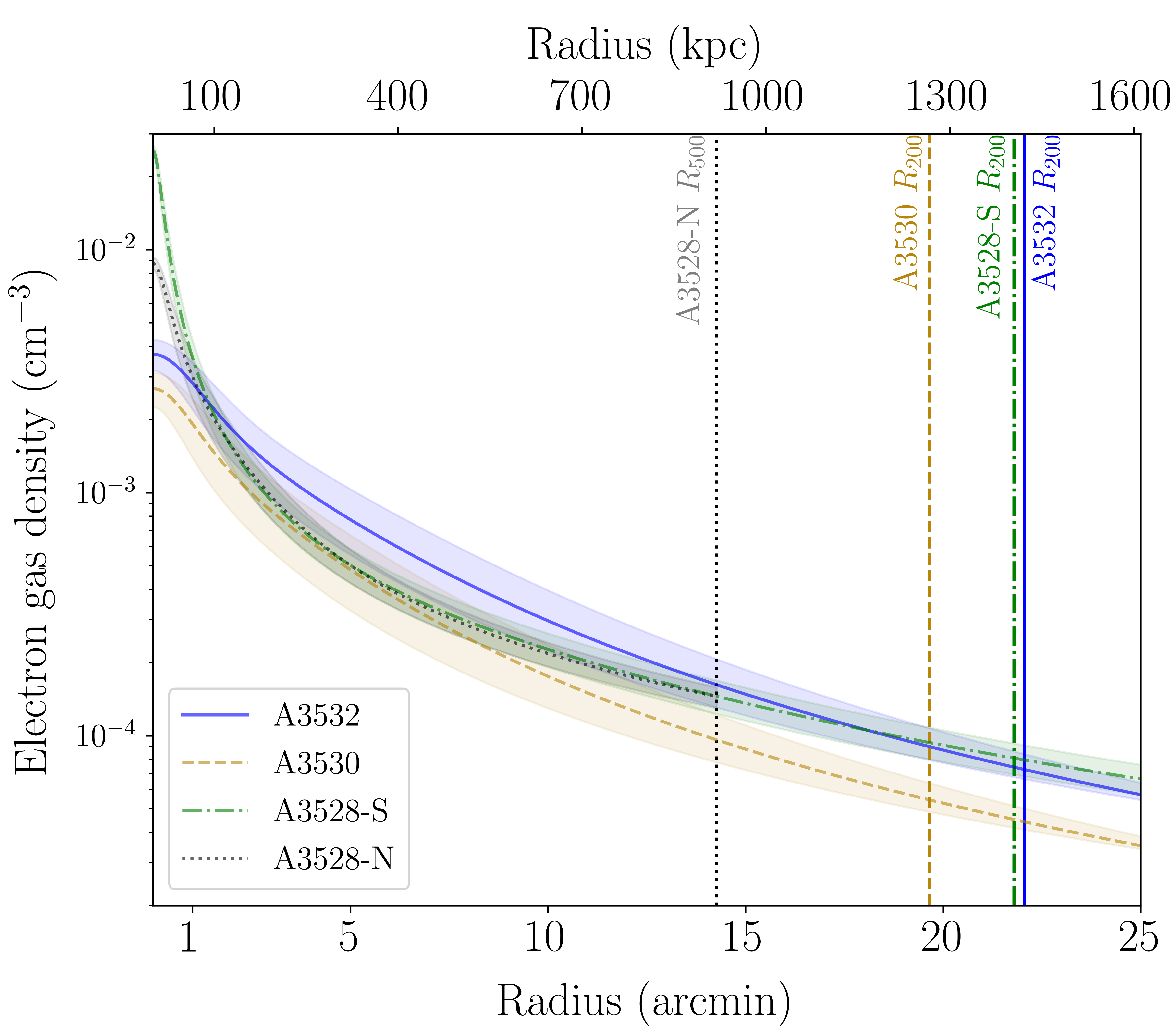}
                \includegraphics[width=0.45\textwidth]{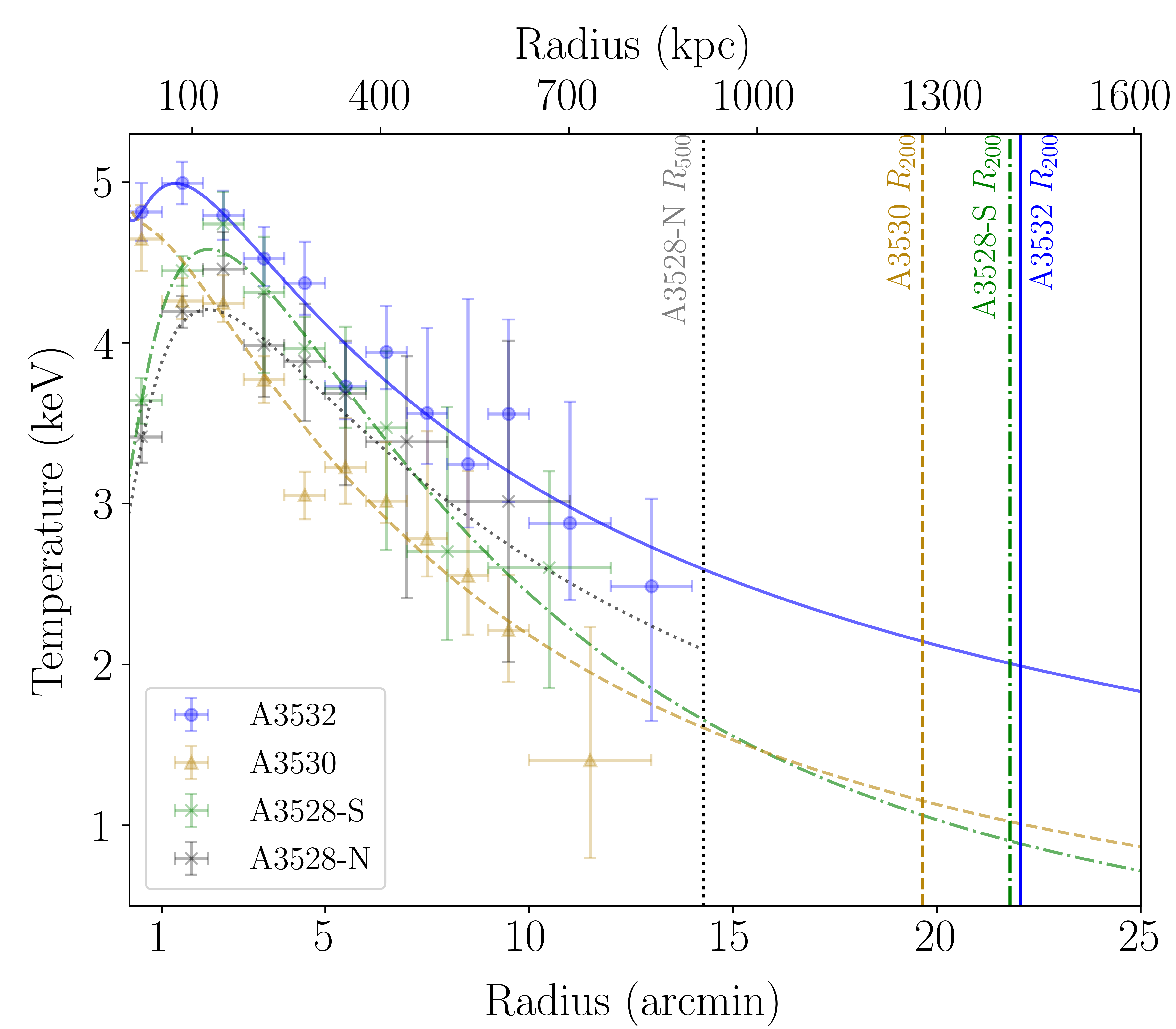}

               \caption{Radial electron gas density (top) and temperature (bottom) profiles for all four clusters. The data points (error bands) of the $n_e$ (\T) profiles are not displayed to avoid overcrowding the figure.}
        \label{ne-T-profiles}
\end{figure}

To derive the SB and $n_e$ profiles of A3532 (A3530), we masked the $R_{200}$ area of A3530 (A3532) and the $>0.5\ R_{500}$ area of A3532 (A3530) that points toward the filament region (i.e., the XMM-Newton pointing with Obs. ID 0861150301). The rest of the available area for each cluster allowed us to derive the SB and $n_e$ profile out to $1.05\ R_{500}$ for A3532 and $1.2\ R_{500}$ for A3532.

To derive the A3528-S profiles, we masked the $R_{500}$ area of A3528-N and the $>0.5\ R_{500}$ area of A3528-S that faces toward the filament (i.e., the XMM-Newton pointing with Obs. ID 0861150201). We exploited the nearby XMM-Newton pointing with Obs. ID 0057740301 (that contains no extended source) and we extracted the SB and $n_\text{e}$ profiles of A3528-S out to $\approx 2\ R_{500}$. Finally, to derive the profiles of A3528-N, we masked the $R_{500}$ area of A3528-S. Due to the limited available area, we could only extract the A3528-N profiles out to $\approx 0.9\ R_{500}$. For each data bin of the A3528-N SB profile, we subtracted the expected SB contamination from A3528-S. The analysis of A3528-N does not have any impact on the X-ray analysis of the filament and it is carried out to simply provide information on this cluster that might be useful for future studies. 

Furthermore, we estimate the X-ray concentration $c_{\text{X}}=\dfrac{L_{\text{X}(<0.1\ R_{500})}}{L_{\text{X}(<R_{500})}}$. The latter is defined as the ratio between the emission originating from the cluster's center ($0.1\ R_{500}$) over the total emission within $R_{500}$. The $c_{\text{X}}$ and central $n_e$ parameters can be used as proxies of the relaxation state of galaxy clusters \citep[e.g.,][]{lovisari17}.

Finally, we extrapolate the best-fit models of A3532 and A3530 out to $2\ R_{500}$, which roughly corresponds to the cluster virial radius $R_{\text{vir}}\approx 2.15 R_{500}$ \citep{reiprich13}. We do so in order to assess the expected contamination of the clusters' signal in every region of the filament. As noted before, for A3528-S we measure the profiles out to this radius. This is crucial if one wishes to robustly isolate any filament emission from the cluster outskirts residual emission. We note that the extrapolation of the A3532 and A3530 profiles to such large radii is susceptible to biases. The expected cluster emission beyond $2\ R_{500}\approx R_{\text{vir}}$, however, is much lower than the CXB and is generally negligible. \citet{lyskova} used tens of stacked SB cluster profiles and showed that generally the cluster emission is at $\approx 2\%$ of the CXB. Thus, mild to moderate offsets between the true and the extrapolated cluster emission at these radial distances should not significantly impact our results.

\subsubsection{Cluster profile behavior}
All SB and $n_e$ profiles are shown in Fig. \ref{SB-profiles} and Fig. \ref{ne-T-profiles}, respectively. All clusters show slightly lower emission than the CXB at $R_{500}$ and significantly lower emission at the cluster limits, that is, $R_{\text{vir}}\approx 2\ R_{500}$. Moreover, all clusters exhibit typical low gas densities ($n_e\lesssim 10^{-4}$) cm$^{-3}$ at $R_{200}$. 

A3532 and A3530 do not show any special features. A3532 is the most luminous system of the four, showing the highest amount of gas out to $\approx 1$ Mpc from the core. A3530 is the less massive cluster, with the lowest $n_e$ at all radii compared to the other systems ($n_e\lesssim 5\times 10^{-5}$ cm$^{-3}$ at $R_{200}$). Both clusters show similar $c_{\text{X}}\approx 0.41-0.44$ and $n_e\approx (3-4)\times 10^{-3}$ cm$^{-3}$ parameters. According to \citet{lovisari17}, these indicate relatively relaxed clusters without a (strong) cool core presence.

On the other hand, the A3528-S/N clusters show strongly centrally peaked emission and gas density, with  $c_{\text{X}}\approx 0.5-0.7$ and $n_e\approx (1-2)\times 10^{-2}$ cm$^{-3}$, respectively. These values indicate highly relaxed clusters \citep{lovisari17} with clear cool core presence. Finally, the emission of A3528-S can be detected out to $R_{200}$, while at $R_{\text{vir}}$, it cannot be distinguished from the CXB. 

\subsection{Temperature profiles}
To construct the \T\ profiles of all clusters, we consider the central bin with a radius of $30^{\prime \prime}$ and all subsequent bins to have a width of $1^{\prime}$. The last two bins of all \T\ profiles have a width of $2^{\prime}$ to obtain sufficient statistics for a robust spectral fit. We extract \T-bins up to the radius for which we have enough sky area and photons to constrain \T. To fit the projected 2D \T\ profiles we use the 3D parametrization presented in \citet{vikh06a} and projected to 2D as explained in \citet{migkas25}. The 3D \T\ profile is given by:
\begin{equation}
    T_{\text{X,3D}}(r) = T_{\text{X,CE}}^* T_{\text{max}} \frac{\left(r/r_{\text{cool}}\right)^{a_{\text{cool}}} + T_0/T_{\text{max}}}{\left(r/r_{\text{cool}}\right)^{a_{\text{cool}}} + 1} \frac{\left(r/r_t\right)^{-a}}{\left(1 + \left(r/r_t\right)^b\right)^{c/b}}
    \label{T-profile-eq}
\end{equation}
where $T_0$, $r_{\text{cool}}$, $a_{\text{cool}}$,$T_{\text{max}}$, and $r_t$, $b$, and $c$ are free parameters. \citet{chen23} constrained the average parameter values of this profile for tens of clusters; however, they showed there is strong variation from cluster to cluster. In this work, the \T\ profiles are constrained to simply obtain a rough estimate of each cluster's \T\ at its outskirts and toward the filament region. This will allow for an approximate comparison of the filament's \T\ (see Sect. \ref{sect:spec-filament-results}) with the expected cluster outskirts \T. Note that the \T\ profile of A3528-N is irrelevant for the study of the filament, but it is presented for completeness.

The \T\ profiles of all clusters are displayed in the bottom panel of Fig. \ref{ne-T-profiles}. The fitted model describes the data well and, if extrapolated, it predicts \T$\lesssim 1$ keV beyond $R_{200}$ for A3530 and A3528-S. A3532 appears to be the hottest cluster at all radii, with \T$\approx 2$ keV at $R_{200}$. However, the extrapolated \T\ profile of A3532 at large radii is marginally biased toward higher values due to the elevated \T\ of the bin at $9.5^\prime$ from the center. When we exclude this bin, the best-fit model predicts \T$\approx 1.2$ keV at $R_{200}$ for A3532. Moreover, fitting the simpler three-parameter \T\ model of \citet{burns10} also describes the data well at large radii, but fails to reproduce the drop in \T\ close to the core for all clusters except A3530. The strongest \T\ drops are found in A3528-N/S, further indicating the presence of a strong cool core in these clusters. Finally, we expect $T\lesssim 0.8$ keV at $R_{\text{vir}}$ for all clusters.

\section{Analysis of the filament}\label{sect:filament-results}

In this section, we present the analysis of the filament connecting the A3528-S cluster with A3532 and A3530, through the analysis of the optical galaxies residing in the filament, the SB profile of the filament's gas, and its spectral characteristics. All these approaches result in a statistically significant detection of a $\approx 7.2$ Mpc filament with \T$\approx (0.8-1.1)$ keV, $n_e\approx (8-10)\times 10^{-6}$ cm$^{-3}$. and a baryonic overdensity of $\delta_{\text{b}}\approx 30-40$ outside the $2\ R_{500}$ of all clusters.

\subsection{Length of the filament in 3D space}

From the analysis of the optical spectroscopic data of the galaxies residing within the Shapley supercluster field, we we determine a filament connecting the clusters A3528-S and A3532/A3530. We identify the spine of the filament and locate its end points situated at the edges of the clusters. The starting point of the filament spine is 254 kpc away from the center of A3528-S, at (RA, DEC)=$(193.61^{\circ}, -29.26^{\circ})$, $4^{\prime}$. The end point of the filament spine is found at (RA, DEC)=$(194.13^{\circ}, -30.30^{\circ})$, close to the intersection of the $R_{500}$ circles of A3532 and A3530.

The projected filament length for the same starting and end points of the spine is found to be $l\approx 4.3$ Mpc. To obtain the deprojected length, we need to estimate the average angle of the filament spine with the plane of the sky. The estimated angle is found to be $\approx 53^{\circ}$ which results in a deprojected length of the filament of $l\approx 7.2$ Mpc. This length categorizes the structure as a short filament according to \citet{galarraga20,galarraga21}.

\subsection{Surface brightness analysis of the filament}\label{sect:sb-filament-results}

To study the axial and radial SB profiles of the filament, we need to unambiguously distinguish any excess filament emission from the cluster outskirts, the CXB, and the residual point source contamination after the masking process. In Fig. \ref{XMM-Suzaku-pointings-mosaic}, one sees that the center of the filament is well outside $2\ R_{500}$ (or $1.3\ R_{200}$) of all clusters where no noticeable cluster emission is expected. However, other regions of the filament partially overlap with the cluster outskirts. Moreover, for Suzaku, a fraction of the AGN photons leak into the unmasked data as described in Sect. \ref{masking_suzaku} which can also overestimate the inferred filament SB. To properly account for the cluster and residual AGN contamination in the filament emission, we utilize a simulated Suzaku count rate image of the cluster-filament complex detailed in Sect. \ref{sect:simul-image-Suzaku}.

\subsubsection{Simulated image to account for cluster and AGN contamination}\label{sect:simul-image-Suzaku}
First, we simulate the emission of A3532, A3530, and A3528-S based on their best-fit SB profiles, extrapolating them out to $2\ R_{500}$ when necessary. For A3528-S we have measured the SB profile out to this radius, so no extrapolation is needed in this case. For A3532 and A3530, we have measured their SB profile to $\approx 1.1\ R_{500}$, but the stable single- or double-$\beta$ functional form of their SB profile allows for a rather safe extrapolation to larger radii, especially since the cluster emission is expected to be $\approx 1\%$ of the CXB level at this distance \citep{lyskova}. XMM-Newton count rate SB profiles are converted to Suzaku count rate SB profiles in \texttt{XSPEC} adopting an absorbed bremsstrahlung emission model (\texttt{tbabs}$\times$\texttt{apec}) assuming the measured core-excised \T\ and $Z$ to be constant throughout each cluster. We also add the uniform CXB emission. Additionally, we simulate all the detected AGN by inducing all their expected Suzaku count rate\footnote{The AGN count rate values were measured by XMM-Newton and converted to Suzaku count rate using an absorbed power law emission model in \texttt{XSPEC} with $\Gamma=1.45$.} into one pixel at their best-fit position, as constrained by XMM-Newton. Then, we convolve the simulated image with the PSF of Suzaku XIS ($1^{\prime}$), which is taken to be constant throughout the XIS FOV. Finally, we apply the same masks as in the real Suzaku data. This method accurately reproduces the residual masked AGN contamination in the filament region due to the poor PSF of Suzaku.

\subsubsection{Results}\label{sect:imaging-filam-results}
Using both the observed and simulated Suzaku images, we measure the radial and axial SB profiles in the box regions displayed in Fig. \ref{XMM-Suzaku-pointings-mosaic}. The box sizes for the radial and axial SB profiles are $35.6'\times 3.9'$ and $4.7'\times 9.7'$, respectively. Since the simulated image accounts for all possible emission contaminants, the observed difference between the two images can be attributed to the filamentary emission. 

\begin{figure*}
\centering
               \includegraphics[width=0.49\textwidth]{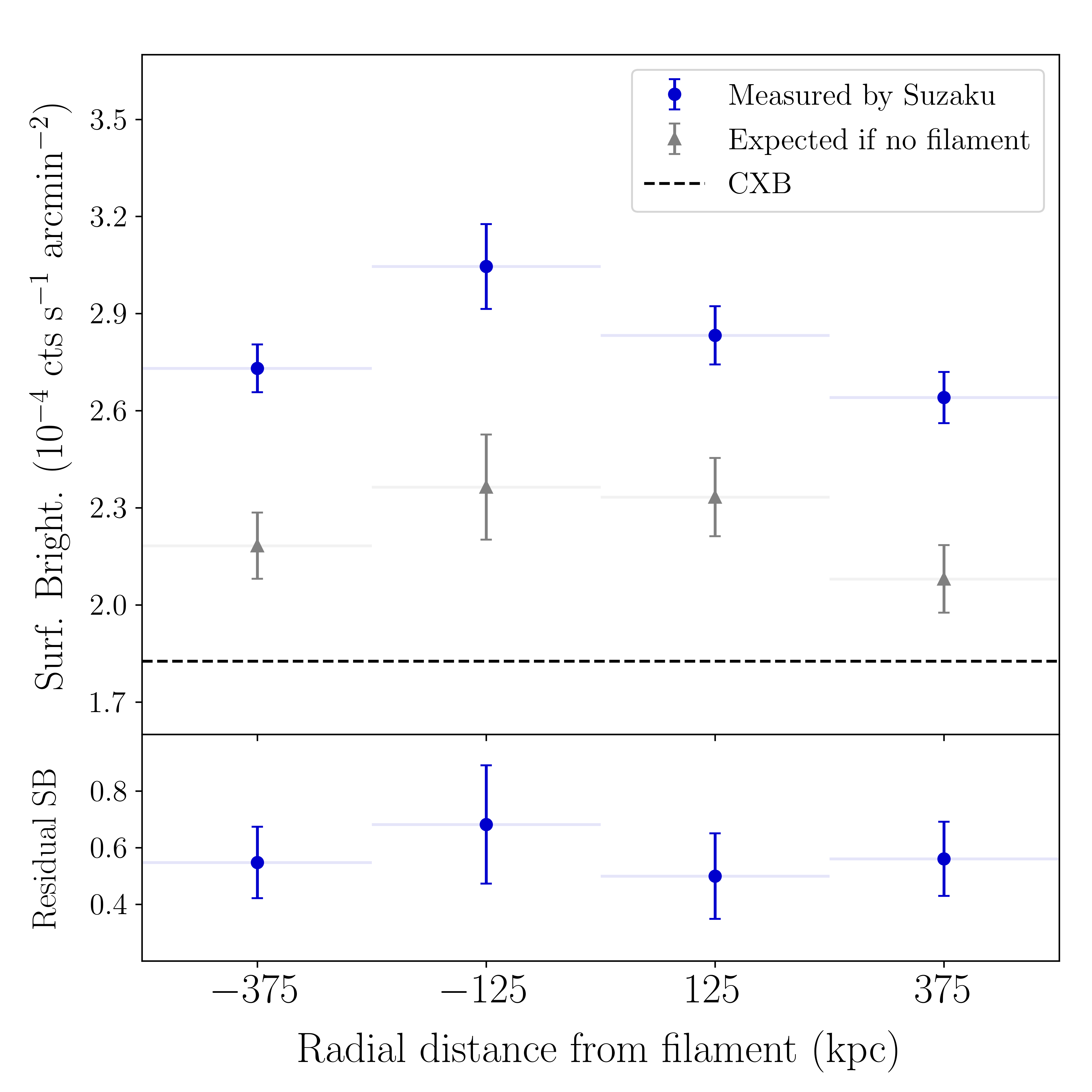}
               \includegraphics[width=0.49\textwidth]{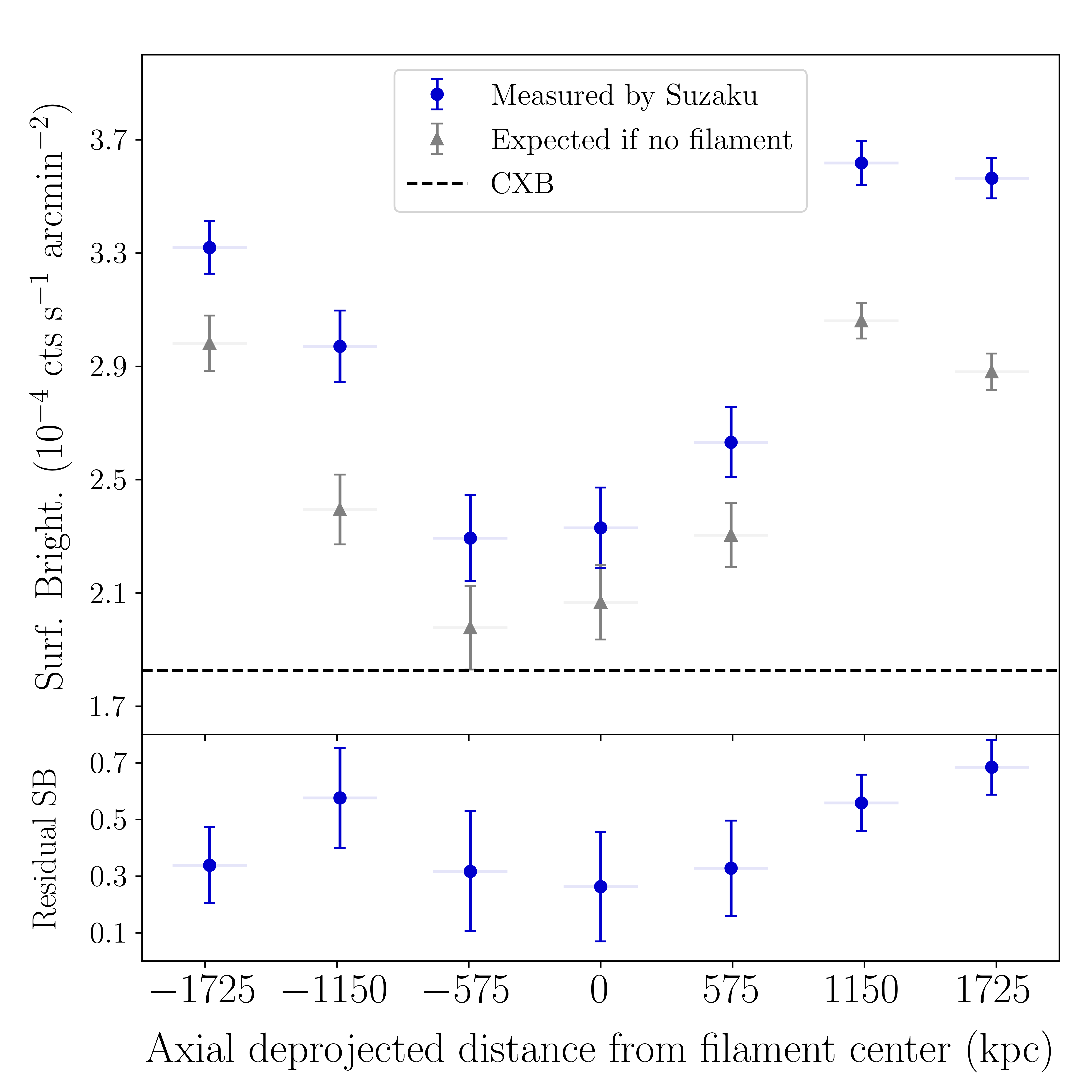}
               \caption{Radial (left) and axial (right) surface brightness profiles of the filament as extracted with Suzaku in the $0.5-2$ keV band. Negative and positive distance values correspond, respectively, to east (south) and west (north) boxes in the left (right) panel of Fig. \ref{XMM-Suzaku-pointings-mosaic}. The blue and gray data points respectively show the measured SB and the expected SB if there was no filament, that is, if all the detected emission came from imperfectly masked AGN, cluster outskirts, and the CXB. The CXB level is displayed with the horizontal dashed black line. In the bottom panels the difference between the measured and expected SB values is shown. These residuals represent the genuine emission coming solely from the filament. The emission of the latter is $\approx (10-30)\%$ higher than the CXB.}
        \label{fig:SB-filament}
\end{figure*}

The results for the radial and axial SB profiles of the filament are shown, respectively, in the left and right panels of Fig. \ref{fig:SB-filament}. We find that the Suzaku data systematically show excess emission compared to the expected SB level if no filament was present. For the radial SB profile, the excess emission is rather constant throughout the four parallel regions and $\pm 375$ kpc from the assumed central filament axis. The lowest and highest excess emission values are $(21\pm 8)\%$ and $(29\pm 11)\%$, respectively, compared to the expected emission from the simulated image without any filament emission, that is, the background\footnote{When we discuss the excess filament emission, we consider the "background" to be the total emission from the CXB plus the residual emission from the imperfectly masked AGN}. Each independent region provides a $\approx (2.5-3)\sigma$ detection of the excess emission. It is important to note that the observed constant filament emission is expected within $\lesssim 600$ kpc from the spine according to cosmological simulations \citep{tuominen21,galarraga22, tuominen23}, thus our result is consistent with simulations. Moreover, the projected spine of the filament (as traced by optical galaxies) oscillates along the axis between the four radial bins; this can potentially result in smoothing out any emission drop away from the spine between the four SB regions. 

For the axial SB profile, excess SB is again detected in all seven independent regions with a total deprojected length of $\approx 4$ Mpc. However, the excess level and its statistical significance varies noticeably. The SB of the filament noticeably decreases toward the central part of the filament's axis, at the maximum distance from all clusters. On the contrary, the SB of the filament peaks at the edges of the connected galaxy clusters. Specifically, the lowest SB excess and statistical significance is found in the central part of the filament, with an excess $(12\pm 10)\%$, detected at $1.2\sigma$. The strongest SB excess, $(24\pm 4)\%$, detected at $6.1\sigma$, is found at the edge of the filament, at the edge of the $R_{200}$ radius of A3528-S. On the other side of the filament, outside the $R_{200}$ areas of A3532 and A3530, the excess SB from the filament is rather noisy, detected at $\approx (2-3)\sigma$. The closest filament region to A3532 and A3530 shows an $(11\pm 5)\%$ SB excess while the next boxed region, $\approx 1.1$ Mpc from the filament's center, shows a $(24\pm 8)\%$ SB excess.

To maximize the signal-to-noise ratio of the excess filament emission compared to the background, we consider the entire region covered by the individual boxes. We find a total SB excess of $(21\pm 3)\%$ compared to the background. This constitutes a $6.8\sigma$ detection of the emission originating from the filament's gas, free of any point source contamination.

\begin{table}
    \centering
    \caption{Best-fit parameter values of the CXB components and the residual emission from imperfectly masked AGN as constrained by the simultaneous fitting of RASS, XMM-Newton, and Suzaku data.}
    \resizebox{\columnwidth}{!}{\begin{tabular}{c c c }
    \hline
    \hline
Component & Parameter & Value\\
    \hline
\texttt{TBabs} &  $N_{\mathrm{H}}$ & ($0.078-0.087$)\\
\hline
\texttt{apec$\mathtt{_1}$} (LHB) & $k_\mathrm{B}T$ [keV] & 0.11\\
& $Z~[Z_\odot]$ & 0.9\\
& $z$ & 0\\
& $norm$ & $9.02\times10^{-6}$\\[5pt]
\hline
\texttt{apec$\mathtt{_2}$} (MWH) &  $k_\mathrm{B}T$ [keV] & 0.19\\
& $Z~[Z_\odot]$ & 0.9 \\
& $z$ & 0\\
& $norm$ & $7.33\times10^{-7}$\\[5pt]
\hline
\texttt{pow}$_1$ & $\Gamma$ & 1.45\\
(unresolved AGN) & $norm$ & $5.32\times10^{-7}$\\[5pt]
\hline
\texttt{pow}$_2$ & $\Gamma$ & 1.45\\
(masked AGN - Suzaku only) & Reg1: $norm$ & $3.46\times10^{-7}$\\[5pt]
& Reg2: $norm$ & $0.71\times10^{-7}$\\[5pt]
& Reg1+2: $norm$ & $2.09\times10^{-7}$\\[5pt]
\hline
    \end{tabular}}
    \label{tab:sky_BG}
\end{table}

\subsection{Spectroscopic analysis of the filament}\label{sect:spec-filament-results}
The spectroscopic analysis of the CXB and the filament regions is carried out following the method described in Sect. \ref{sect:xmm-spectral} and \ref{spectral-analysis-suzaku}, respectively. The best-fit values of the CXB model components are given in Table \ref{tab:sky_BG}, while the XMM-Newton CXB spectra with their best-fit model are shown in the bottom panel of Fig. \ref{fig:spec-fit-filament}. The spectral best-fit values of the filament emission are shown in Table \ref{tab:spectral-fits} and the Suzaku spectra of Regions 1 and 2 are displayed in the top and middle panels of Fig. \ref{fig:spec-fit-filament}, respectively.

\subsubsection{Sky background and imperfectly masked AGN contribution}
The LHB (\texttt{apec$_1$}), MWH (\texttt{apec$_2$}), and unresolved AGN (\texttt{pow$_1$}) components show rather typical values according to our experience \citep[e.g.,][]{migkas20}. Interestingly, the (\texttt{pow$_2$}) component, describing the residual contamination by masked AGN in the Suzaku data, is comparable to the \texttt{pow$_1$} component for Region 1 (as defined in Sect. \ref{spectral-analysis-suzaku}), suggesting that an imperfect AGN masking can have a considerable impact on the spectral fits if not properly taken into account. On the other hand, the \texttt{pow$_2$} value is much lower for Region 2 than for Region 1, and than (\texttt{pow$_1$}). This is partially expected due to the presence of the bright (masked) AGN in Region 1 at (RA, DEC)=$(193.917^{\circ}, -29.610^{\circ})$. We note that the \texttt{pow$_1$} and \texttt{pow$_2$} components are not strongly degenerate due to the tight constraints of \texttt{pow$_1$} from XMM-Newton, independent of \texttt{pow$_2$}. The \texttt{pow$_2$} component is also not strongly degenerated with the emission from the filament (\texttt{apec$_3$}), since the latter is almost negligible at $\gtrsim 3$ keV, where \texttt{pow$_2$} dominates.

\subsubsection{WHIM emission from the filament}

\begin{figure}[htbp]
\centering
        \includegraphics[width=0.45\textwidth]{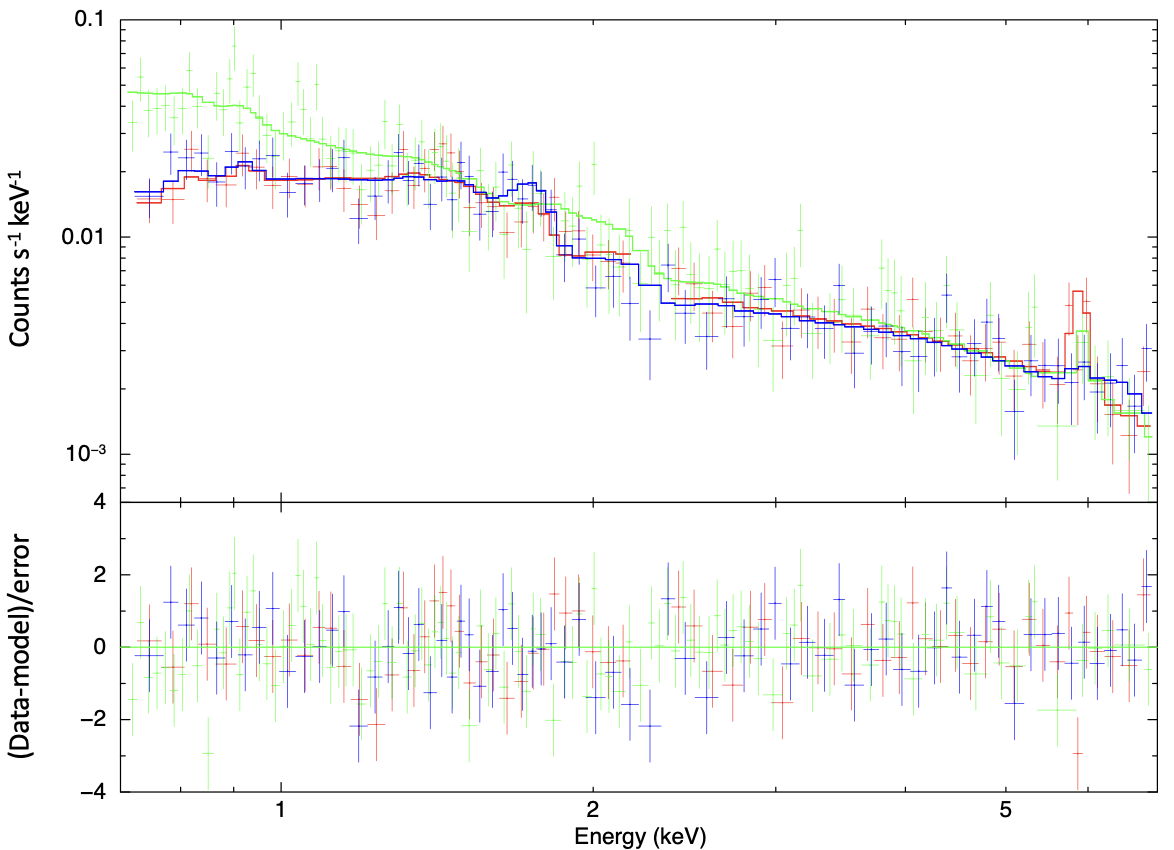}
        \includegraphics[width=0.45\textwidth]{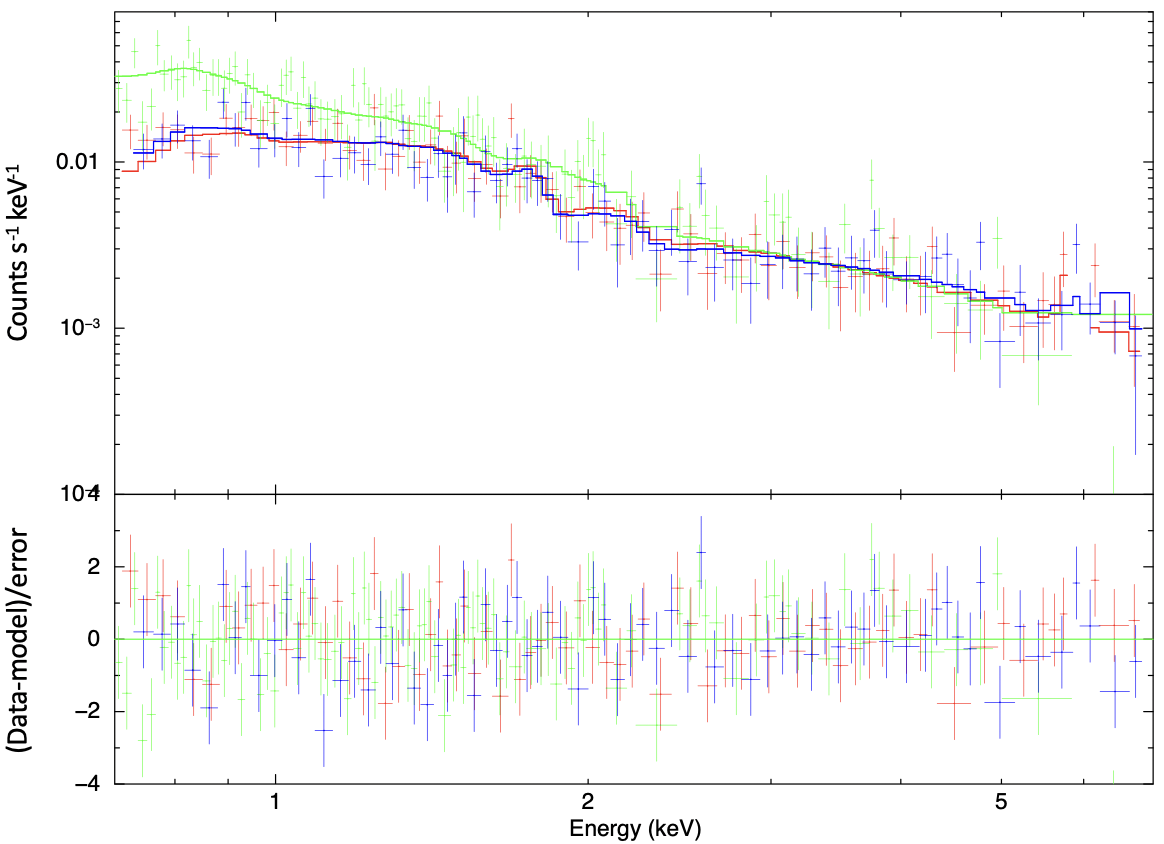}
        \includegraphics[width=0.45\textwidth]{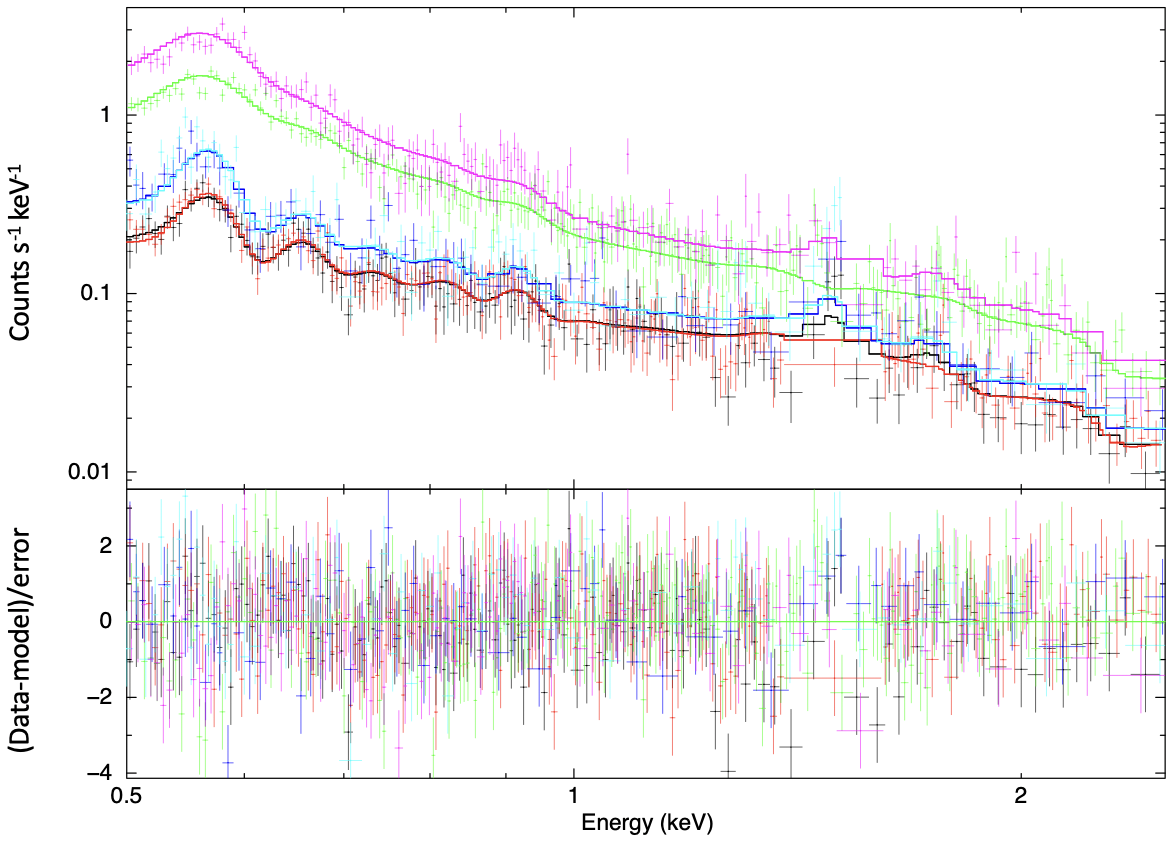}. 
        \caption{Spectra of the central filament regions extracted with Suzaku (top and middle) for XIS0 (blue), XIS1 (green), and XIS3 (red), and for the CXB regions extracted with XMM-Newton (bottom). The convolved best-fit models are also plotted for each detector. The residuals of the data compared to the model are displayed in the bottom of each figure.}
        \label{fig:spec-fit-filament}
\end{figure}

The Suzaku spectra from Regions 1 and 2, together with their best-fit model and fit residuals, are displayed in the top and middle panels of Fig. \ref{fig:spec-fit-filament}. The best-fit model parameters are shown in Table \ref{tab:spectral-fits}. Employing 1000 mock spectra sets as described in Sect. \ref{spectral-analysis-suzaku}, we find that the model provides an excellent fit for both Regions with $p-$value$\approx (0.3-0.6)$, or a $<1\sigma$ agreement between the model and the data.

Both Regions 1 and 2 show a positive \texttt{apec$_3$} normalization, supporting the existence of WHIM emission at $2\sigma$. The $norm$ value is similar for both regions, with $1.82_{-0.90}^{+0.55}\times 10^{-6}$ cm$^{-5}$arcmin$^{-2}$ and $2.61_{-1.29}^{+0.91}\times 10^{-6}$ cm$^{-5}$arcmin$^{-2}$ for Regions 1 and 2, respectively. The $norm$ values are rather low, making the \texttt{apec$_3$} component comparable to the total background model only in the $\approx (0.7-2)$ keV range, after which the summed AGN emission dominates. Both Region 1 and 2 show a low \T\ plasma, with $0.85_{-0.34}^{+0.53}$ keV and $1.09_{-0.39}^{+0.31}$ keV, respectively. The filament's \T\ is close to the value expected for the outskirts of the surrounding clusters, although the sources are far apart.  Furthermore, we estimate the electron density, $n_e$, of the filament based on the $norm$ values. To do so, we assume that the filament has a cylinder shape with uniform gas density, a radius of 1.4 Mpc (similar to the $R_{200}$ of the three connected galaxy clusters)\footnote{This choice is supported by \citet{galarraga21} and \citet{tuominen21} that find the filament density to increase up to $\approx 1$ Mpc from the spine}, height and width of the emission volume equal to the boxes from which we extracted the spectra, and a $53^{\circ}$ inclination of the filament, following \citet{dietl24} and \citet{veronica24}. For Regions 1 and 2 we find $7.99_{-2.47}^{+1.80}\times 10^{-6}$ cm$^{-3}$ and $9.56_{-3.08}^{+2.12}\times 10^{-6}$ cm$^{-3}$, respectively. To compute the baryon overdensity $\delta_{\text{b}}$ of the filament, we adopt a normalized baryon density of $\Omega_{\text{b}}=0.056$ \citep{planck20} and consider the critical density of the Universe at $z=0.054$ for the $\Lambda$CDM cosmology we use. We find that Regions 1 and 2 have $\delta_{\text{b}}=31^{+6}_{-10}$ and $\delta_{\text{b}}=37^{+6}_{-10}$, respectively. This constitutes the first X-ray detection of a filament with such a low baryonic overdensity. This indicates that the studied filament is a pristine one, and our analysis successfully removes most of the contribution from residual AGN emission and galaxy halos scattered in the filament that might result in an overestimation of $\delta_{\text{b}}$. Finally, we also estimate the electron pressure of the filament gas, $P_e$, finding $\approx (7-10)\times 10^{-6}$ keV cm$^{-3}$, slightly lower than typical $P_e$ values from cluster outskirts.

Because Regions 1 and 2 show consistent results for the properties of the WHIM, we combine the spectra from the two regions and fit them with a single \texttt{apec$_3$} model component to maximize the S/N ratio, labeling the joint region as Region 1+2. The combined fit also returns a satisfyingly good fit (Table \ref{tab:spectral-fits}). For Region 1+2, we find $norm=2.51_{-0.81}^{+0.53}\times 10^{-6}$ cm$^{-5}$arcmin$^{-2}$, $T=0.93_{-0.32}^{+0.28}$ keV, $n_e=9.56_{-3.08}^{+2.12}\times 10^{-6}$ cm$^{-3}$, and $\delta_{\text{b}}=36^{+7}_{-8}$. Consequently, WHIM is detected at $3.1\sigma$ based on our spectroscopic analysis of the purely thermal excess emission of the filamentary gas. Finally, by considering the volume of the filament and assuming a uniform $n_e$, we estimate its total gas mass to be $M_{\text{gas}}=1.18\times 10^{13}\ M_{\odot}$, which is comparable to the $M_{\text{gas}}$ of the four galaxy clusters connected by the filament. 

\begin{table*}
    \centering
    \caption{Best-fit values for the spectral fitting of the different filament regions outside $2\ R_{500}$ of all clusters.}
    \resizebox{\textwidth}{!}{\begin{tabular}{c c c c c c c c}
    \hline
Source & $norm$ & $k_\mathrm{B}T$ & $Z$ & $n_\mathrm{e}$ & $P_{\mathrm{e}}$ & $\delta_{\text{b}}$ & C-stat/d.o.f./$p-$value\\
(\texttt{apec$\mathtt{_3}$}) & $[10^{-6}$ cm$^{-5}/$arcmin$^2]$ & $[$keV$]$ & $[Z_{\odot}]$ & $[10^{-6}$ cm$^{-3}]$ & $[10^{-6}$ keV cm$^{-3}]$ & \\[5pt]
\hline \\%[-1.7ex]
Region 1 & $1.82_{-0.90}^{+0.55}$ & $0.85_{-0.34}^{+0.53}$ & $0.1^{+0.05}_{-0.05}$ & $7.99_{-2.47}^{+1.80}$ & $6.79_{-3.43}^{+4.50}$ & $31^{+6}_{-10}$ & 219/256/0.61 \\[5pt]
\hline \\[-1.7ex]
Region 2 & $2.61_{-1.29}^{+0.92}$ & $1.09_{-0.39}^{+0.31}$ & $0.1^{+0.05}_{-0.05}$ & $9.56_{-3.08}^{+2.12}$ & $10.42_{-5.02}^{+3.76}$ & $37^{+6}_{-10}$ & 344/309/0.32 \\[5pt]
\hline \\[-1.7ex]
Region 1+2 & $2.51_{-0.81}^{+0.53}$ & $0.93_{-0.32}^{+0.28}$ & $0.1^{+0.05}_{-0.05}$ & $9.38_{-1.44}^{+0.82}$ & $8.72_{-3.29}^{+2.72}$ & $36_{-8}^{+7}$ & 581/565/0.44\\[5pt]
\hline
\hline
    \end{tabular}}
    \label{tab:spectral-fits}
    \tablefoot{The metallicity $Z$ was fixed to $0.1\ Z_{\odot}$ for the default results and was varied between $(0.05-0.15)$ $Z_{\odot}$ to account for the systematic uncertainty induced by this assumption.}
\end{table*}

\section{Discussion}\label{discussion}
Spectroscopic detection of WHIM emission in X-rays from individual filaments, independently of any cluster outskirt emission, is especially rare. A very few, recent studies that used eROSITA data simultaneously characterized $n_e$ and $T$ for the gas residing in cosmic filaments, confirming the existence of WHIM baryons \citep{dietl24, veronica24}.These studies both reported higher-than-expected baryon overdensities, however, which typically disagree with simulations (see Sect. \ref{past_studies}). Large $\delta_{\text{b}}$ values might indicate that residual emission from unresolved AGN and galaxy halos (or distant galaxy groups), or shocks due to gas compression during ongoing cluster mergers, is mixed with the WHIM emission, resulting in an overestimation of $n_e$ and $\delta_{\text{b}}$. In this work, we report the first-ever X-ray spectroscopic detection of WHIM emission originating from a $\delta_{\text{b}}\lesssim 50$ filament, suggesting a pristine filament.

\subsection{Consistency between the imaging and spectroscopic analyses of the filament}

We detected WHIM X-ray emission coming solely from the filament at $6.1\sigma$ and $3.1\sigma$ through SB and spectroscopic analyses, respectively. To check the robustness of our results, we need to examine the consistency of the two results, that is, if the level of the excess filament emission compared to the background agrees between the two methods. To do so, we use the full model presented in Eq. \ref{eq:spectral_model_suzaku} and the best-fit parameter values in Table \ref{tab:spectral-fits} to compute the emission of the filament and the background in the energy band used in the imaging analysis, that is, $0.5-2$ keV. We find that, in that band, the $\mathtt{tbabs\times apec_3}$ component for Region 1+2 returns an absorbed flux of $9.53\times 10^{-16}$ erg cm$^{-2}$ s$^-1$ while the background (i.e., the sum of the rest of the model components for the same area as Region 1+2) returns $3.17\times 10^{-15}$ erg cm$^{-2}$ s$^-1$. Consequently, the WHIM emission shows an $\approx 30\%$ excess compared to the background in the spectroscopic analysis. Given the uncertainties of the model parameters, this result is consistent with the results of the SB analysis, where the excess emission from the entire filament was $(21\pm 3)\% $. This highlights the robustness of our analysis thanks to the detailed method we followed.

\subsection{Cross-calibration differences between Suzaku and XMM-Newton}\label{cross-calib}

Past studies have shown that several X-ray instruments exhibit cross-calibration differences between them that result in discrepant measured properties of X-ray sources, for instance, different galaxy cluster \T\ and power law indexes of AGN \citep[e.g.,][]{nevalainen,schellenberger,wallbank,migkas24}. In our study, we combined XMM-Newton and Suzaku data to jointly assess the CXB contribution to the filament's SB signal and spectra, and detect the excess WHIM emission. For the spectral analysis, we used XMM-Newton data to quantify the contribution of the LHB ($\mathtt{apec_1}$) and MW ($\mathtt{apec_2}$) emission to the soft part of the filament's Suzaku spectra ($\lesssim 1.5$ keV) and the contribution of unresolved AGN ($\mathtt{pow_1}$) to the harder part of the filament's Suzaku spectra ($\gtrsim 2$ keV).\footnote{The emission level of all these components was obtained by jointly fitting Suzaku and XMM-Newton data. However, the latter dominate the fit due to the higher count statistics.} During the joint fitting process of Suzaku and XMM-Newton data, we assumed no significant calibration differences between XMM-Newton and Suzaku, for the following reasons. \citet{kettula} showed that when comparing the measured \T\ of the same clusters with both instruments, no significant differences were found in the cross-calibration of the EPIC/PN camera of XMM-Newton and the XIS detectors of Suzaku for the $0.5-7$ keV band (we use the $0.7-7$ keV band). Moreover, they used the stacked residual method to show that no significant cross-calibration differences ($\lesssim 10\%$) are observed in the $0.8-2$ keV band, especially for soft X-ray sources such as clusters with $T\lesssim 3$ keV. All these suggest that linking the $norm$ values of the $\mathtt{apec_1}$ and $\mathtt{apec_2}$ model components between XMM-Newton and Suzaku, important only within $0.7-1.5$ keV, is a safe assumption that does not considerably impact our results. The same holds for the $\mathtt{pow_1}$ at soft energies. Similarly, for the SB analysis of the filament, XMM-Newton data are utilized to determine the total flux within the $0.5-2$ keV band of the CXB and compare it with the SB from the filament region obtained by Suzaku. Based on the results from \citet{kettula}, no systematic flux difference is expected between the two telescopes in this band.

At harder energies, data from the two instruments are combined to constrain the total AGN emission, from unresolved ($\mathtt{pow_1}$, XMM-Newton) and resolved ($\mathtt{pow_2}$, Suzaku) point sources. The $norm$ value of $\mathtt{pow_2}$ is allowed to vary within $\pm 30\%$ from the value that returns the measured residual flux of the imperfectly masked AGN (see Sect. \ref{spectral-analysis-suzaku}). The value of $\mathtt{pow_2}$ determined by Suzaku strongly depends on the independent constraint of $\mathtt{pow_1}$ from XMM-Newton. \citet{kettula} showed that there is good agreement in the derived cluster \T\ in the hard band ($2-7$ keV) between EPIC/PN and XIS. Moreover, \citet{tsujimoto11} and \citet{madsen17} showed that EPIC/MOS and XIS return similar $\Gamma$ and fluxes for AGN, within $\lesssim 10\%$. Consequently, the $\pm 30\%$ variation range of the $norm$ value of $\mathtt{pow_2}$ would allow the component to shift accordingly to cancel out any mild cross-calibration differences of the $\mathtt{pow_1}$ component between XMM-Newton and Suzaku. We are not directly interested in the absolute value of $\mathtt{pow_2}$; instead, we focus on the \emph{sum} of $\mathtt{pow_1}+\mathtt{pow_2}$ that affects the emission of the filament. Therefore, we believe our results are not significantly impacted by cross-calibration differences between XMM-Newton and Suzaku and our modeling method is sufficient to account for any such mild differences. Nevertheless, when one is jointly using data from different telescopes to analyze diffuse, low SB sources, unknown cross-calibration issues might increase the systematic uncertainties of the derived results. Thus, future work needs to further explore the effects of (even mild) cross-calibration differences between individual X-ray telescopes when searching for WHIM emission.

, \subsection{Comparison with simulations and previous studies}\label{past_studies}

The expected thermodynamic properties of cosmic filaments have been studied using cosmological, hydrodynamical, large-scale structure simulations. \citet{martizzi19} used the TNG100 Illustris simulations \citep{spirgel18} to define WHIM as gas with $n_e\lesssim 10^{-4}$ cm$^{-3}$ and $T\lesssim 1$ keV (their Fig. 4), with hotter gas labeled as hot medium and denser gas labeled as warm circumgalactic medium. According to this categorization, our analysis supports the existence of gas lying at the threshold between WHIM and the hot medium. \citet{galarraga21} used the TNG300-1 simulation box of IllustrisTNG and found that the thermodynamic properties of the filament we study in this work ($n_e\approx 10^{-5}$ cm$^{-3}$ and $T\approx 1$ keV), are common for filaments in the local Universe, while they are not as common for collapsed halos (their Fig. 4). The filament's gas properties we derive in this work lie again between WHIM and hot gas. Moreover, the $P_e$ value of the studied filament in this work, is slightly higher, but marginally consistent, with the $P_e$ values close to the central axis of short filaments ($\leq$ 9 Mpc length), as found in \citet{galarraga21}. Furthermore, the $n_e$ and $\delta_{\text{b}}$ values we find for our filament are highly consistent with the values found close to the central axis of short filaments by \citet[][Fig. 2 and 4]{galarraga22} and in filaments within the EAGLE simulations \citep{tuominen21}. This strongly suggests that the emission we detect in this work originates genuinely from the WHIM in a short, pristine filament, without any significant contamination from scattered, unresolved AGN throughout the filament, galactic haloes, or emission from galaxy groups and clusters. This is likely due to the sophisticated spectroscopic analysis method we followed to account for all kinds of residual emission unrelated to the filament's gas. Finally, in Fig. \ref{fig:SB-filament} we do not see any trend of the excess emission within 375 kpc from the central filament axis. However, this is to be expected, because the radial density profiles of filaments only start to drop noticeably at larger radii from the central axis, according to \citet{galarraga22}.

Contrary to our results, similar previous studies have found significantly higher $n_e$ and $\delta_{\text{b}}$. \citet{dietl24} and \citet{veronica24} both found $n_e\approx5\times 10^{-5}$ cm$^{-3}$ and $\delta_{\text{b}}\approx 200$ for the filaments they detected using eROSITA data (from the All-Sky Survey and PV phase, respectively). Such gas (over)densities typically correspond to virialized halos. As \citet{dietl24} noted, their measured filament $n_e$ might be boosted due to unresolved emission from collapsed halos, small groups, gas clumps, and even AGN, throughout the filament. This is particularly likely given the shallow eROSITA data used, with slightly worse spatial resolution, which might prevent the robust detection of such contaminating sources. In the case of \citet{veronica24}, contamination from the outskirts ($\approx R_{200}$) of the surrounding clusters is also likely to contaminate the filament's emission. Moreover, \citet{zhang24} used deeper, stacked eROSITA data of 7817 filaments identified with optical data and carefully estimated the contamination caused from unmasked sources to the emission of the filaments. Doing so, they constrained the average filament baryon overdensity to be $\delta_{\text{b}}=76^{+38}_{-26}$, consistent with out findings within $1.5\sigma$. However, their detection still suggests that filaments are twice as dense as the filament we probed and as suggested by \citet{galarraga22}.

Nevertheless, all three past studies found $T\approx (0.7-1.1)$ keV for the gas of their studied filaments, which is consistent with our results. Moreover, the derived \T\ of the WHIM in all three studies, and our own, is close to the \T\ values expected at cluster outskirts. This might suggest the existence of an unknown physical mechanism (not identified by simulations) that heats the gas in filaments above a certain threshold or commonly shared limitations to detect cooler WHIM, or overestimating its true \T.

\subsection{Implications for our understanding of the large-scale structure}

The observational detection of filaments and WHIM emission significantly alleviates the missing baryons problem. Locating where a large fraction of cosmic baryons reside helps us to accurately map the large-scale structure and compare it with the predictions of $\Lambda$CDM and cosmological simulations. Moreover, the detection and analysis of filaments is crucial to accurately trace the mass distribution in the local Universe. Better mapping the cosmic mass distribution can provide useful information about the peculiar velocity field and its consistency with $\Lambda$CDM. Recent studies suggest that coherent flow motions (bulk flows) of galaxies and clusters are much larger in amplitude and scale than expected in the standard cosmological model \citep{migkas21, watkins23}. Such bulk flows require large mass concentrations in the $z\lesssim 0.1-0.2$ Universe that have not been identified yet, as concentrations such as the Shapley supercluster are believed to not contain enough mass to justify the observed bulk flows. Until now, the mass of filaments was ignored since the latter remained undetected. If, however, more cosmic filaments start to be directly detected and it is found that their mass contribution is non-negligible, this might partially explain the amplitude of the observed bulk flows and reduce the tension with $\Lambda$CDM. 

Moreover, our result partially alleviates the tension between past eROSITA studies \citep{dietl24,veronica24} and cosmological simulations \citep[e.g.,][]{galarraga21,tuominen21} regarding $n_{\text{e}}$ and $\delta_{\text{b}}$ values of WHIM in filaments. However, we also find a rather higher-than-expected $T$, similar to past studies, but this is subject to large statistical uncertainties.

\subsection{The crucial role of XMM-Newton for robustly treating AGN emission}

To accurately study the diffuse, low SB X-ray emission of the filament, it is crucial to properly distinguish the emission of point sources residing in the cosmic filament from the WHIM emission. To do so, we utilized deep XMM-Newton pointings to identify AGN, mask them, and estimate their residual contamination in the Suzaku spectra from the filament region. This residual emission from resolved AGN by XMM-Newton is then modeled during the spectral fits with the $\mathtt{pow_2}$ component in Eq. \ref{eq:spectral_model_suzaku}. To further evaluate the robustness of this method, we estimated the sum of the masked AGN flux, the residual flux from imperfectly masked AGN as estimated from $\mathtt{pow_2}$ model component, and the flux from the $\mathtt{pow_1}$ component from the Region 1+2. This sum corresponds to the total flux of resolved and unresolved AGN in the studied sky region. We compared this flux sum with the integrated AGN flux from the CDFS, down to $10^{-16}$ (far below the detection limit of XMM-Newton for the available exposure times). We found that the two values impressively agree within $\approx (10-15)\%$. This highlights that the novel applied method for treating AGN contamination robustly isolates the WHIM emission from AGN photons. 

However, not all studies that searched for WHIM emission in filaments used deep, high-resolution X-ray data. Shallower exposure times or worse angular resolution of the X-ray data might result in higher levels of contamination from AGN and an overestimation of $n_{\text{e}}$ and \T\ of the filamentary gas. To assess the impact of the lack of deep, high-resolution X-ray data in the search for diffuse WHIM emission, we repeat our analysis using only the Suzaku data to treat AGN emission. The CXB modeling remains unchanged, based on XMM-Newton data. We use $2^{\prime}$ radius masks for all detected AGN with Suzaku as in the default analysis (see Sect. \ref{masking_suzaku}) and ignore all AGN not resolved by Suzaku. We then repeat the spectral fits for Region 1+2. We find that $\mathtt{pow_2}$ increases by $\approx 85\%$, the density, temperature, and baryon overdensity of the filament also increase. Specifically, we find $T=1.81_{-0.28}^{+0.33}$ keV, $n_e=1.87_{-0.39}^{+0.46}\times 10^{-5}$ cm$^{-3}$, and $\delta_{\text{b}}=78^{+19}_{-18}$. Even with the use of Suzaku data only, the WHIM emission is still significantly detected. However, the thermodynamic properties of the filament gas are overestimated. Most importantly, \T\ increases by a factor of two, resembling the \T\ of gas at clusters' outskirts. Moreover, while $\delta_{\text{b}}$ shifts closer to the results of past studies as discussed in Sect. \ref{past_studies}, it remains rather low. These results suggest that, in the presence of considerable unresolved AGN emission, modeling the latter might not be sufficient for obtaining unbiased results. Even though detection of filament emission might still be possible with low-resolution (or shallow) X-ray data, deep data with high angular resolution are necessary to accurately treat the emission of point sources and obtain an unbiased characterization of WHIM emission.

\section{Summary}

The detection of WHIM emission from cosmic filaments is essential for alleviating the missing-baryon problem and for better understanding the large-scale structure. However, very few studies have reported an X-ray detection of the emission that originates from individual filaments, and even fewer studies have analyzed the WHIM emission spectrally. In this work, we reported the unambiguous X-ray detection of a newly discovered 7.2 Mpc long cosmic filament by imaging and spectroscopic analysis. The filament was recently discovered through its optical galaxy overdensity by \citet{aghanim2024} and shows an inclination of $\approx 53^{\circ}$ to the plane of the sky. It connects two galaxy cluster pairs in the Shapley supercluster, namely the A3532/30 and A3528-N/S cluster pairs. To properly account for any residual cluster emission when studying the filament, we used XMM-Newton data to fully characterize the four galaxy clusters through their surface brightness, temperature, and gas density profiles. The four clusters have intermediate masses of $\approx 2\times 10^{14}\ M_{\odot}$, with A3528-N/S showing signs of a strong cool core and high relaxation, while A3530/32 appear to have non-cool cores, but are also relatively relaxed. By exploiting the high angular resolution of XMM-Newton, the sensitivity of Suzaku to diffuse, low SB sources, the excellent sky coverage of the entire cluster-filament complex and their surroundings by the two telescopes, and an innovative imaging analysis, we studied the axial and radial SB profiles of the filament region. We detected $\approx (10-30)\%$ excess X-ray emission compared to the background from different filament regions at a $6.1\sigma$ level. 

To study the spectra from filament regions free of any cluster emission, we employed deep Suzaku data and a sophisticated method to fully account for the emission of AGN residing in the filament. We utilized XMM-Newton to detect and characterize the AGN and to measure the sky background around the cluster-filament complex. Through the spectral analysis, we detected WHIM emission in two independent regions of the filament. Overall, we measured the temperature and gas density of the central part of the filament to be $T\approx 0.9$ keV and $n_e\approx 10^{-5}$ cm$^{-3}$, respectively. The total gas mass of the filament is $\approx 1.2\times 10^{13}\ M_{\odot}$. The baryon overdensity of the filament is $\delta_{\text{b}} \approx 36$. This constitutes the first detection of such a low-density single filament without stacking techniques. Our findings agree well with the thermodynamic properties of filaments as predicted by cosmological simulations of the large-scale structure. On the other hand, previous studies have reported $\approx 5$ times higher $\delta_{\text{b}}$ for other filaments. Our detailed methodology for effectively removing contamination from AGN emission allowed us to trace the WHIM emission that purely originates from the pristine filament we studied, while previous studies might have been more affected by residual emission from halos and point sources. This may have caused them to overestimate the density of the filamentary gas. Finally, we showed that when only Suzaku data are used (which do not resolve all other sources throughout the filament), the gas properties are significantly affected. This biases the final results. Consequently, high-quality deep X-ray data and a detailed modeling of all types of X-ray emission in the studied sky area are crucial for a robust characterization of the WHIM.

\begin{acknowledgements}
We thank the anonymous referee for their useful comments that improved our manuscript. We thank Thomas Reiprich for useful discussions about the Suzaku data analysis and spectral fitting. This study was based on observations obtained by XMM-Newton, an ESA science mission with instruments and contributions directly funded by ESA member states and the USA (NASA). K.M. acknowledges support in the form of the X-ray Oort Fellowship at Leiden Observatory. 
K.M. and F.P. acknowledge support from German Federal Ministry for Economic Affairs and Energy (BMWi) provided  through DLR under project 50OR2101. N.A. and T.T. acknowledge support from the European Union’s Horizon 2020 research and innovation program grant agreement ERC-2015-AdG 695561 (ByoPiC project) and from the grant agreement ANR-21-CE31-0019 / 490702358 from the French Agence Nationale de la Recherche / DFG (LOCALIZATION project). TT also acknowledges the support of the Academy of Finland grant no. 339127.
\end{acknowledgements}

\bibliographystyle{aa} 
\bibliography{XXX}          

\appendix
\section{Combined image highlighting the emission of the filament}

\begin{figure}[hbtp]
\centering
               \includegraphics[width=0.5\textwidth]{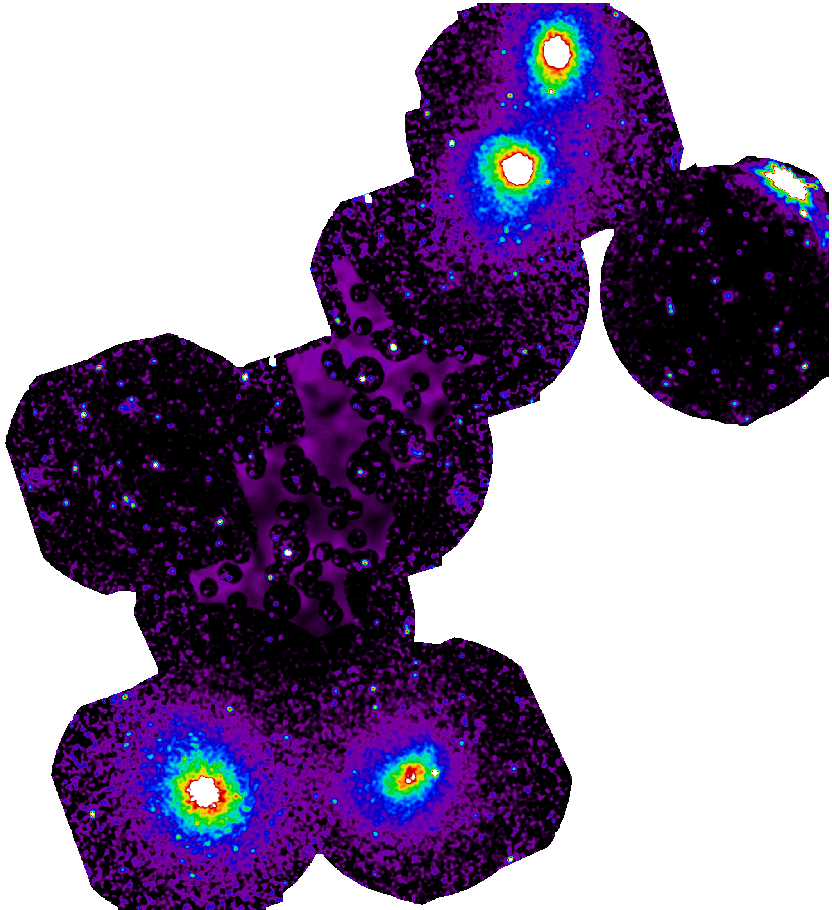}

               \caption{Fully cleaned (as in Fig. \ref{XMM-Suzaku-pointings-mosaic}), CXB-subtracted, combined XMM-Newton-Suzaku mosaic image in the $0.5-2$ keV band. The Suzaku data were calibrated to the XMM-Newton/MOS count rate. The emission from the cluster outskirts and the AGN residuals was removed from the Suzaku images, thus what is shown here is pure filamentary emission as detected by Suzaku. At the Suzaku's masked areas, the underlying XMM-Newton image is shown instead.}
        \label{fig:Combined-mosaic-image}
\end{figure}

To better visualize the X-ray emission from the cosmic filament, we created a combined XMM-Newton–Suzaku image by overlaying Suzaku’s detection of the filament onto the XMM-Newton mosaic. First, we removed residual emission from cluster outskirts and imperfectly masked AGN in the Suzaku data by subtracting the simulated image described in Sect. \ref{sect:simul-image-Suzaku} from the Suzaku count-rate mosaic shown in the right panel of Fig. \ref{XMM-Suzaku-pointings-mosaic}. Next, we calibrated the Suzaku count rates to those of XMM-Newton/MOS using the instruments' response files and the best-fit spectral model of the filament plus the CXB in \texttt{XSPEC}, accounting for differences in pixel size. This yielded an image of the filament's intrinsic emission, free from contaminants and expressed in units equivalent to the XMM-Newton/MOS count rate. We then projected this clean Suzaku image onto the XMM-Newton mosaic in the left panel of Fig. \ref{XMM-Suzaku-pointings-mosaic}. In regions where the Suzaku data were masked due to contamination by AGN, we display the unmasked XMM-Newton image to reveal the bright AGN behind each mask. Finally, we subtracted the average CXB level and smoothed the image. The color scale is adjusted to enhance the visibility of the filament, whose emission is only $\approx$21\% of the CXB. The resulting image is shown in Fig. \ref{fig:Combined-mosaic-image}

\end{document}